# Mechanical Signatures of Liquid Polymorphism in Glass–Glass Transitions

*Yizhi Liu[1], Yixiang Chen[1], Shaoqin Jiang[1], Xun-Li Wang[2*], Gang Sun[3*], and Limei Xu[1,4,5*]*

*1 International Center for Quantum Materials, School of Physics, Peking University, Beijing 100871, China*

*2 Department of Physics and Center for Neutron Scattering, CityUHK, 83 Tat Chee Avenue, Kowloon, Hong Kong, China*

*3 Center for Advanced Quantum Studies, School of Physics and Astronomy, Beijing Normal University, Beijing 100875, China*

*4 Interdisciplinary Institute of Light-Element Quantum Materials and Research Center, Peking University, Beijing 100871, China*

*5 Collaborative Innovation Center of Quantum Matter, Beijing, China*

## Abstract

Liquid–liquid phase transitions and critical phenomena have been proposed in many supercooled liquids, yet direct access to them is often prevented by vitrification. This raises the question of whether signatures of the underlying liquid-state criticality can persist after kinetic arrest. Here, using a family of waterlike isotropic models, we find that glass-glass transition (GGT) sharpness correlates with the stability of the liquid–liquid critical point (LLCP) relative to the glass transition. Within the potential energy landscape, low- and high-density liquid and glass states are organized within the same two-megabasin landscape. The GGT pathways follow the mechanical stability limits of amorphous basins associated with the parent liquid. When $T_g < T_c$, the LLCP lies above the glass transition, and the GGT crosses a mechanically unstable region, producing a van der Waals–like loop in the inherent structure pressure. When $T_g > T_c$, vitrification occurs before the critical region is reached. The same two megabasins remain, but the GGT proceeds through a smooth structural transformation without an abrupt mechanical instability. Across the tunable model family, the mechanical response of the GGT varies systematically with LLCP stability, providing a quantitative connection between liquid-state polymorphism and glassy behavior. This result indicates that the mechanical character of an experimentally accessible GGT may therefore provide an indirect readout of liquid behavior otherwise hidden by vitrification.

## Introduction

Equilibrium thermodynamics provides a rigorous framework for phase transitions in ergodic systems, but glasses are nonergodic: they are path dependent, retain long-lived memory, and relax slowly and nonexponentially(1). As Anderson emphasized, this makes the glass transition a central unresolved problem in condensed matter physics(2). Parisi's work provided a landmark theoretical framework for spin glasses and a broader language for complex disordered systems(3), yet a unified quantitative microscopic description of structural glasses remains incomplete. When a liquid is cooled below the glass transition temperature $T_g$, it becomes an amorphous solid that preserves short- and medium-range structural order while acquiring rigidity(4-12). Beyond vitrification, some amorphous materials undergo glass-glass transitions (GGTs), in which one amorphous state transforms into another, often between low- and high-density amorphous forms(13-21). Representative examples include water, gallium, and metallic glasses (8, 14-17, 19, 20, 22, 23). These GGTs display remarkable diversity, ranging from abrupt first-order-like jumps to smooth crossovers. What determines this variation remains unclear.

A possible clue to this variation comes from liquid polymorphism. GGTs bear similarities to the proposed liquid–liquid phase transitions (LLPTs) between distinct liquid states in the deeply supercooled regime(24-31). In particular, both can exhibit sharp, first-order-like transitions(14, 15) and comparable density contrasts. A recent simulation also shows that glasses formed near an LLCP can retain enhanced long-range density fluctuations (6). Moreover, studies have linked the pressure dependence of glass-transition and heating-induced GGT loci to the presence of an LLPT(32, 33). These parallels suggest a possible underlying connection between glassy transitions and liquid-state criticality. Establishing whether a liquid–liquid critical point exists is important for understanding liquid polymorphism, but direct experimental access to this region is often prevented by rapid crystallization(34-37). GGTs could provide a practical approach from the low-temperature glassy regime to examine the liquid–liquid transition hypothesis experimentally, if their mechanical character is strongly and robustly correlated with liquid-state criticality. This raises a sharper question: do GGTs merely resemble LLPTs phenomenologically, or do they preserve a mechanistic imprint of liquid-state criticality?

Within the potential energy landscape (PEL) framework, both liquid polymorphism and amorphous polyamorphism can be described in terms of inherent structures organized into basins and megabasins(38, 39). Studies of aging glasses showed that nonequilibrium configurations can explore basin families associated with equilibrium liquids at an effective temperature(7, 40). Previous PEL studies of SPC/E water, ST2 water, a waterlike monatomic model, and TIP4P/2005 water showed that glass transformations follow pathways distinct from those sampled by the equilibrium liquid(41-44). Except for SPC/E water, these models also exhibit anomalous $P_{IS}(v)$ behavior during the GGT, including a van der Waals–like loop. These studies establish several forms of liquid–glass correspondence in the PEL, but do not identify what controls the mechanical stability of the GGT pathway between LDA and HDA.

Here, we address this problem using a continuously tunable Fermi–Jagla (FJ) model family(45). We show that the GGT is mechanically connected to the parent liquid, with the transition tracing the stability limits of the corresponding liquid-derived ISs. Within the PEL framework, a mechanical instability occurs as the glass approaches the stability limit of an amorphous basin. This connection suggests that whether the parent liquid reaches the critical region before vitrification is associated with whether the arrested glass subsequently undergoes such a mechanical instability during the GGT. Across the FJ model family, varying the interaction softness systematically shifts the relative positions of the glass-transition temperature $T_g$ and the liquid–liquid critical temperature $T_c$. When $T_c > T_g$, the LLCP remains accessible before vitrification and the subsequent GGT is first-order-like; when $T_c < T_g$, kinetic arrest occurs first and the GGT becomes smooth. The mechanical character of the GGT therefore reflects LLCP stability.

## Results

### Stability limits of liquid-derived inherent structures

We perform molecular dynamics (MD) simulations of the isotropic Fermi–Jagla (FJ) model(45), a waterlike two-length-scale model that exhibits both liquid and glass polymorphism. We first consider the $A_0 = 4.7$ model, where $A_0$ controls the softness of the interaction potential (Methods). In this model, the LLCP is identified from the highest-temperature crossing of equilibrium liquid isochores in the P-T plane (Methods), which gives $T_c$ and $P_c$. To determine $T_g$, we cool an equilibrated liquid from $T = 0.50\epsilon_0/k_B$ to $T = 0.01\epsilon_0/k_B$ isobarically at $P_c$. Because $T_g$ depends on the cooling rate, we use the slowest cooling rate explored, giving $T_g = 0.117\epsilon_0/k_B$ (Fig. S1; Methods). For this model, the liquid–liquid critical point (LLCP) is accessible before vitrification ($T_c > T_g$, as shown in Fig. 1a).

To compare the liquid-liquid phase transition (LLPT) and glass–glass transition (GGT) within a unified framework, we perform energy minimizations on configurations sampled along both transitions. The resulting local minima of the potential energy landscape (PEL) are called inherent structures (ISs) (Methods). We characterize each IS by $P_{IS}(v)$, the pressure of the minimized configuration at specific volume $v$. For the GGT, $P_{IS}(v)$ is averaged over ten independent compression and decompression trajectories between LDA and HDA at $T = 0.01\epsilon_0/k_B$ (Methods). For the LLPT, $P_{IS}(v)$ is calculated from energy-minimized ISs obtained from 100 statistically independent equilibrium liquid configurations sampled at $T = 1.1\mathrm{T}_g$ for each volume (Fig. 1b). Consistent with earlier work on this model(44), we find that the LLPT and the GGT show similar PEL signatures, most notably a van der Waals–like loop in $P_{IS}(v)$ (Fig. 1b). Unlike the LLPT, which is an equilibrium first-order transition, the GGT is a mechanically driven nonequilibrium transformation. Why the two exhibit similar van der Waals–like loops in $P_{IS}(v)$ therefore remains unclear.

To examine the origin of this similarity, we further analyze the mechanical stability of these liquid-derived ISs (black symbols in Fig. 1b and Fig. 1c). At each volume, ten independent liquid configurations are energy-minimized and subsequently compressed and decompressed at the same temperature used for the GGT ($T = 0.01\epsilon_0/k_B \approx 0.1T_g$). Near their initial volumes, $P_{IS}(v)$ changes approximately linearly (Fig. 1c). In the low-pressure (large-volume) regime, the mean-square displacement (MSD) remains small and changes smoothly (Fig. 1d), while the bulk modulus $B = -v(\partial P/\partial v)$ remains positive (Fig. 1e). These observations indicate that the liquid ISs remain mechanically stable and deform elastically within their initial basins.

Upon further compression or decompression, the liquid-derived ISs associated with the unstable, positive-slope region of the van der Waals–like loop in $P_{IS}(v)$ (Fig. 1c) can no longer remain mechanically stable. They undergo pronounced nonelastic deformation and collapse onto the corresponding GGT branch (Fig. 1c). Under compression, the transition points (blue squares in Fig. 1c) trace the LDA-to-HDA branch of the conventional GGT loop on the potential energy landscape (PEL); under decompression, the corresponding points (purple squares in Fig. 1c) trace the HDA-to-LDA branch. At these transition points, $P_{IS}(v)$ shows a positive slope and the MSD increases abruptly during both compression (Fig. 1d) and decompression (Fig. S2). In addition, the bulk modulus $B$ decreases and reaches a minimum close to zero (Fig. 1e). By contrast, liquid-derived ISs originating from the stable regions of the liquid, where $P_{IS}(v)$ has a negative slope, remain mechanically stable. ISs derived from the HDL region undergo only elastic deformation upon compression, whereas those derived from the LDL region remain elastic upon decompression. This behavior indicates that the mechanical stability of the parent liquid is inherited by the corresponding liquid-derived ISs.

To further characterize the transition, we examine the relative IS energy, $\Delta E(v) = E_{IS}(v) - E_{IS,liquid}(v)$ (Fig. 1f). Here, $E_{IS}(v)$ is traced along a compression or decompression path of liquid-derived ISs. $E_{IS,liquid}(v)$ is the fitted function (Fig. S3cd) of the mean IS energy of 100 independent liquid configurations sampled at $T = 1.1T_g$ for each volume. Near each reference state, $\Delta E(v)$ shows a quadratic volume dependence (harmonic-like) on volume during elastic deformation. As the trajectory approaches the transition, $\Delta E(v)$ departs from this quadratic behavior.

Together, these observations identify strong mechanical softening accompanied by structural rearrangement. The corresponding transformation points trace the conventional GGT branches, supporting a mechanical connection between the GGT and the response of liquid-derived ISs. Therefore, the van der Waals–like loop in the GGT branch can be viewed as a series of stability-limit points of liquid-derived ISs prepared at different volumes and is related to the mechanical stability of these liquid-derived ISs. This suggests that vitrification preserves not only structural features of the parent liquid, but also its mechanical stability. When the liquid–liquid critical region is accessible before vitrification, these mechanical features may be retained in the glass, giving rise to a first-order-like GGT.

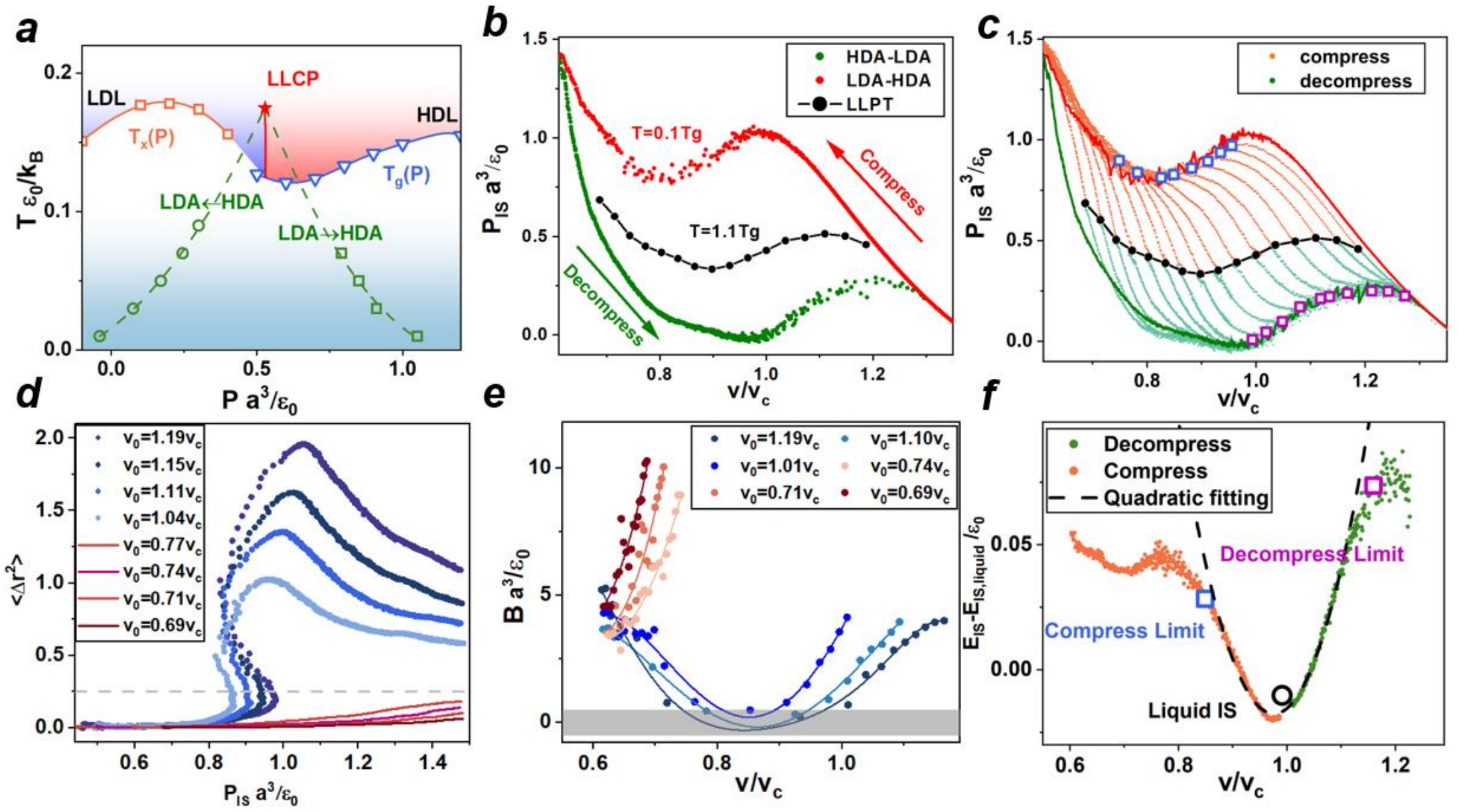

**Figure 1. Stability limits of liquid-derived ISs.** a, Temperature–pressure T-P phase diagram of the FJ model with softness parameter $A_0$=4.7, showing the liquid–liquid critical point (LLCP) at $T_c$=$0.175\epsilon_0/k_B$. Glass-transition temperatures (blue triangles) and crystallization temperatures (orange squares) are obtained at different pressures using the same cooling rate. Blue and orange lines are guides to the eye. Glass–glass transition pressures are indicated by green symbols, with circles marking the transition upon decompression and squares marking the transition upon compression; green dashed lines are guides to the eye. b, Inherent-structure pressure $P_{\mathrm{IS}}$ across the liquid–liquid phase transition (LLPT) at $T = 1.1T_g$ and the glass–glass transition (GGT) at $T = 0.01\epsilon_0/k_B \approx 0.1T_g$. For the GGT path, the initial low-density amorphous (LDA) state is prepared by rapidly quenching an equilibrated low-density liquid at $T = 1.1T_g$ $(P < P_c)$ and is subsequently compressed and decompressed (Methods). Volume is normalized by the critical volume, $v_c \approx 2.85a^3$. c, $P_{\mathrm{IS}}$ along compression and decompression paths initiated from liquid-derived ISs at different initial volumes. At each volume, ten temporally separated equilibrium liquid configurations are energy-minimized to generate the liquid-derived ISs (filled black symbols). The conventional GGT path from b is shown for comparison. Blue squares mark the transition points along compression, and purple open squares

mark those along decompression. These points are operationally defined by the fixed MSD thresholds (dashed line) specified in the Methods and lie near the positions of the maximum MSD increase along the corresponding paths. They trace the LDA-to-HDA and HDA-to-LDA branches, respectively, of the conventional GGT loop. d, MSD along the compression paths. Circle-marked curves denote paths initiated from ISs derived from LDL configurations, whereas solid curves denote paths initiated from ISs derived from HDL configurations. e, Bulk modulus $B = -v(\partial P/\partial v)$ obtained from the pressure response $P(v)$. The bulk modulus decreases and reaches a minimum close to zero in the GGT region. f, Relative inherent-structure energy $\Delta E(v) = E_{IS}(v) - E_{IS,\mathrm{liquid}}(v)$ during compression and decompression paths. Here, $E_{IS}(v)$ is traced along each path, and $E_{IS,\mathrm{liquid}}(v)$ is the mean energy of 100 liquid-derived ISs prepared at the same volume (Methods). Near each initial state, $\Delta E(v)$ follows the quadratic volume dependence expected for harmonic elastic deformation. Departure from this behavior indicates the breakdown of harmonic elasticity.

## A shared landscape for liquid and glassy transformations

To characterize the energetic organization of the PEL underlying the GGT and LLPT, we examine the IS energy $E_{IS}(v)$ sampled along both transitions. In the FJ model, $E_{IS}(v)$ depends primarily on volume and remains nearly monotonic through the transition (Fig. S3ab). The LLPT and GGT are mainly reflected in the emergence of negative curvature in $E_{IS}(v)$ (Inset in Fig. S3ab), consistent with previous work(44). Because the absolute value of $E_{IS}(v)$ contains this volume-dependent contribution, it does not directly reveal the relative stability of amorphous states along a transition pathway involving different volumes. Rather than the absolute energy, we are interested in how stable an IS is relative to the lowest-energy reference state available at a given volume. We therefore reference each IS to the crystalline state at the same volume and define $\Delta E_{IS}(v) = E_{IS}(v) - E_{IS,crystal}(v)$ (Methods). $E_{IS,crystal}(v)$ is obtained by interpolating the lowest sampled crystal IS energy at different volumes (Fig. S4, Methods). This quantity measures the excess energy of each amorphous IS relative to the crystalline reference at the same volume. A relaxed glass typically occupies relatively low-energy amorphous minima, whereas strongly compressed or distorted configurations can lie at substantially higher excess energies. The liquid samples a broader set of configurations, whose average excess energy can be compared with the glass on the same basis. Referencing to the crystal removes much of the direct volume dependence and allows liquid- and glass-derived ISs to be compared directly.

Using the excess energy $\Delta E_{IS}(v)$, we map the sampled inherent-structure landscape in the $(P_{IS}, v)$ plane. The ISs are obtained from equilibrium liquid configurations at $T = 1.1T_g$ and from their subsequent compression and decompression trajectories. For visualization, we divide the $(P_{IS}, v)$ plane into small bins and average the $\Delta E_{IS}(v)$ of all ISs within each bin (Methods). The resulting landscape shows two distinct regions corresponding to the low-density (LDL/LDA) and high-density (HDL/HDA) megabasins, separated by a ridge of higher excess energy (red dashed line in Fig. 2a). Within this landscape, the LLPT follows a lower-excess-energy pathway between the two megabasins, whereas the GGT follows a higher-excess-energy nonequilibrium pathway. To examine the temperature dependence of the GGT pathways, we performed LDA–HDA and HDA–LDA transformations at several temperatures below $T_g$, using the same compression and decompression rates. As the temperature increases toward $T_g$, the GGT pathways move closer to the LLPT pathway (Fig. 2a). This behavior is consistent with faster structural relaxation at higher temperatures, which allows the glass to explore lower-energy configurations. However, the GGT remains a nonequilibrium transformation because of kinetic arrest.

Beyond the transformation temperature, the cooling rate of LDA may also influence the GGT pathways. We prepared LDA states by isobaric cooling at $P < P_c$ using four different cooling rates, $q_T = 10^{0}\text{-}10^{-3}\,\epsilon_0{}^{3/2}/m^{1/2}k_B a$. For each cooling rate, ten independent LDA configurations were generated and subsequently subjected to compression–decompression cycles at $T = 0.01 \approx 0.1T_g$. Changing the cooling rate produces distinct glassy initial states, with faster cooling producing higher-energy, lower-volume states (Fig. 2b). Upon subsequent compression and decompression, these glasses follow trajectories that closely resemble those of liquid-derived ISs obtained at similar volumes. This correspondence shows that different cooling histories produce different glassy states, while their subsequent GGT pathways are closely related to the corresponding liquid-derived IS trajectories.

At a fixed volume, the interaction potential and volume constraint define the PEL, and the ISs correspond to its local minima. Among the amorphous states, the LDA- and HDA-derived ISs occupy the low-energy regions of the corresponding low- and high-density basins. Although the $E_{IS} - P_{IS}$ plots do not represent the full multidimensional landscape, they provide fixed-volume views (Fig. 2c,d) that support the two-megabasin picture in Fig. 2a. The equilibrium LDL and HDL states lie within the corresponding low- and high-density basins, consistent with the two-megabasin landscape in Fig. 2a. At the same volumes, the compression and decompression states lie near different edges of these basins, reflecting the distinct nonequilibrium pathways followed during the GGT. The relation to the crystalline state also differs between the two sides. At low density, the crystalline states lie near the bottom of the same basin occupied by the LDL/LDA-derived ISs (Fig. 2c), suggesting a low-energy route toward crystallization. At high density, the HDL/HDA-derived ISs and crystalline states occupy distinct basins separated by higher-energy configurations (Fig. 2d), indicating a larger energetic barrier between the amorphous and crystalline states.

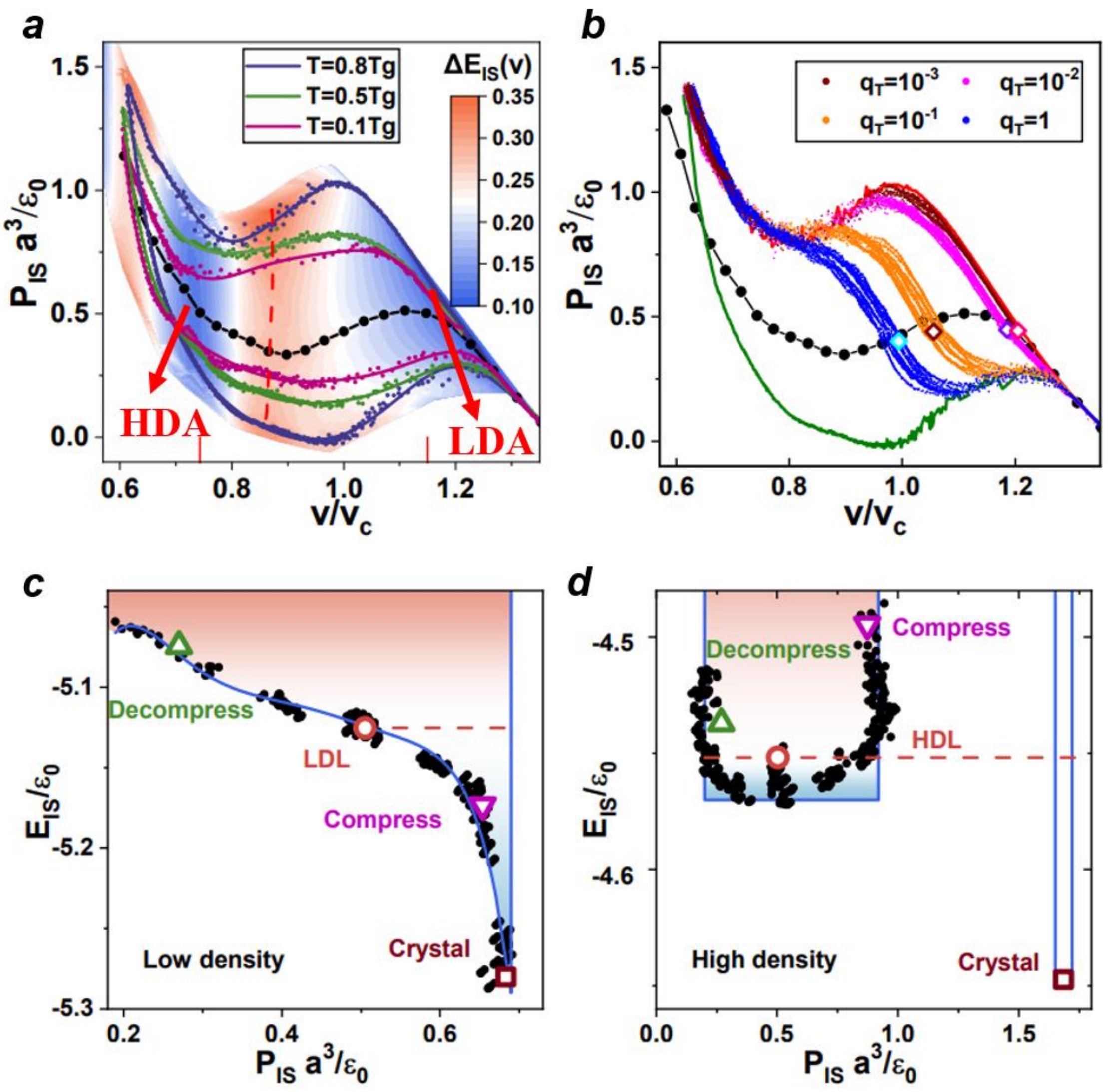

**Figure 2. Energy landscape and kinetic pathways of the GGT and LLPT.** a, Inherent-structure energy relative to the crystalline reference, $\Delta E_{IS}(v) = E_{IS}(v) - E_{IS,crystal}(v)$, for the FJ model with $A_0$ = 4.7, shown as a function of inherent-structure pressure $P_{IS}$ and specific volume $v$. Colors represent the average $\Delta E_{IS}(v)$ of the ISs within each $(P_{IS}, v)$ bin. The map reveals two distinct megabasins associated with low- and high-density amorphous states, separated by a higher-energy ridge (dashed line). b, Cooling-rate dependence of the GGT. Open diamonds indicate glassy initial states prepared at low pressure ($P < P_C$) using cooling rates from $q_T = 10^0$ to $10^{-3}$ $\epsilon_0{}^{3/2}/m^{1/2}k_B a$, followed by isothermal compression and decompression at $T = 0.01\epsilon_0/k_B$. Ten independent trajectories are included for each cooling rate. Colors identify the cooling rates indicated in the legend. Red and green curves denote the reference GGT in Fig.1. c, Inherent-structure energy $E_{IS}$ as a function of inherent-structure pressure $P_{IS}$ at a low-density specific volume ($v$=1.15$v_C$), along the red line indicated in panel a. d, Same as c, but at a high-density specific volume ($v$=0.74$v_C$). Volume is normalized by the critical volume, $v_c \approx 2.85a^3$. In c and d, maroon circles denote

ISs sampled from equilibrium liquids, green and violet triangles denote ISs sampled along the HDA→LDA and LDA→HDA pathways, respectively, and dark-red squares denote crystalline states.

## LLCP stability and the mechanical character of glass–glass transitions

For the FJ model with soft interaction ($A_0 = 4.7$), the transition points of liquid-derived ISs trace the conventional GGT path. Next, we examine whether this correspondence persists when the LLCP is buried below $T_g$. Here, $A_0$ controls the softness of the interaction potential, with $A_0 = 4.7$ and 6.6 corresponding to softer and harder interactions, respectively. For $A_0 = 6.6$, the temperature–pressure phase diagram is obtained from equilibrium MD simulations (Fig. 3a). No LLCP is observed in the equilibrated liquid regime. However, the equilibrium isochores approach one another as the temperature decreases (Fig. S5d). By extrapolating these isochores below $T_g$, we estimate the location of the LLCP ($T_c < T_g$) (Methods; Fig. S5cd).

Next, we repeat the analysis of liquid-derived ISs for ($A_0 = 6.6$) (Fig. 3b). The transition points of the liquid-derived ISs again converge onto the conventional GGT path (Fig. 3b). Unlike the $A_0 = 4.7$ model, the equilibrium-liquid response, the GGT path, and the locus traced by the transition points of liquid-derived ISs are all smooth and monotonic, with no van der Waals–like loop. When $T_c < T_g$, the two low- and high-density megabasins remain present in the IS landscape (Fig. 3c) and the GGT still connects the low- and high-density megabasins, but proceeds smoothly without the abrupt mechanical instability observed when $T_c > T_g$ ($A_0 = 4.7$, Fig.2a). The low- and high-density megabasins exhibit shapes similar to those for $A_0 = 4.7$ (Fig. S6). As the temperature increases toward $T_g$, the GGT pathways approach the $T = 1.1T_g$ liquid pathway but remain smooth and monotonic over the explored glassy temperature range. Consistent with the monotonic $P(v)$ response, changing the cooling rate has only a weak effect on the initial LDA state (Fig. S7).

We then extend this comparison across the FJ model family by systematically varying the interaction softness ($A_0 = 4.7, 5.0, 5.4, 5.8$ and 6.6). For soft interactions $A_0 = 4.7 - 5.4$ ($T_c > T_g$), the LLCP remains accessible before vitrification (Fig. 3d). The subsequent GGT is first-order-like, with a pronounced van der Waals–like loop in $P_{IS}(v)$. For hard interactions $A_0 = 5.8 - 6.6$ ($T_c < T_g$), kinetic arrest occurs before the liquid reaches the critical region. The corresponding GGT is smooth and mechanically stable. We therefore distinguish an LLCP-stable regime ($T_c > T_g$) from a kinetically buried regime ($T_c < T_g$). For the intermediate models, the stability-limit points of liquid-derived ISs also trace the corresponding GGT paths (Fig. S8). The liquid- and glass-derived $P_{IS}(v)$ responses evolve in parallel as the interaction becomes harder (Fig. 3e). The van der Waals–like loops in both the liquid and GGT progressively weaken and eventually become smooth and monotonic. Thus, as LLCP stability decreases, the two low- and high-density megabasins remain, but the GGT between them changes from a first-order-like transition with an abrupt mechanical instability to a smooth transformation.

To quantify this relation, we use the maximum slope of the inherent-structure equation of state, $(dP_{IS}/dv)_{max}$, to characterize the mechanical response of the GGT. Positive values indicate a mechanically unstable region along the IS pathway, while negative values indicate a mechanically stable transformation (Fig. S9). We use the normalized temperature separation, $(T_c - T_g)/T_c$, as an operational measure of LLCP stability: positive values correspond to $T_c > T_g$, and negative values to $T_c < T_g$. Across the FJ family, $(dP_{IS}/dv)_{max}$ increases systematically with $(T_c - T_g)/T_c$, linking greater LLCP stability to a stronger mechanical instability of the subsequent GGT (Fig. 3f). A linear fit to the FJ data gives a slope of $k_{FJ} = 1.76$. This relation shows that, within the tunable FJ family, the mechanical character of the GGT reflects whether the parent liquid critical region was stable before vitrification.

This correlation observed across the FJ family raises the question of whether a similar trend holds for other polyamorphic systems. In the case of water, we digitized published $P_{IS}(v)$ at the lowest possible temperature for the SPC/E, ST2, and TIP4P/2005 water models (41-43) (Fig. S10a-c). The corresponding values of $(dP_{IS}/dv)_{max}$ are -0.07261, 0.1135, and 0.0737, respectively. The positive values for ST2 and TIP4P/2005 are consistent with more pronounced, first-order-like GGT signatures, whereas the negative value for SPC/E is consistent with a smooth, continuous transition. We further compare these signatures with the relative positions of the liquid–liquid critical temperature $T_c$ and glass-transition temperature $T_g$, using estimates from previous studies (6, 25, 46-48). For ST2 water, we adopted the reported LLCP temperature $T_c = 246\,K$ at $P_c = 0.208\,GPa$ (46) and estimated $T_g(P_c) \sim 241.4\,K$ from the reported kinetic-arrest boundary (46). These values give $(T_c - T_g)/T_c \sim 0.019$. For TIP4P/2005 water, we adopted the reported LLCP at $T_c = 172K$ and $P_c = 0.1861GPa$ (25), together with $T_g \sim 170K$ reported for isobaric cooling near the critical pressure (6). These values give $(T_c - T_g)/T_c \sim 0.0116$.

For SPC/E water, the LLCP temperature has been estimated to lie in the range $T_c \sim 140 - 175\,K$ based on a potential-energy-landscape (PEL) analysis (47). Giovambattista et al. reported $T_g \sim 188\,K$ from the heat-capacity response upon heating glass configurations prepared by cooling at constant volume (48). Although this estimate does not correspond to the isobaric cooling protocol used for the other models, it provides a useful reference for comparison. These estimates yield $-0.343 < (T_c - T_g)/T_c < -0.074$. As an independent thermodynamic comparison, in the harmonic-Gaussian PEL phase diagram of Sciortino et al. (47), the predicted LLCP lies below the Kauzmann locus, in a region where the model yields negative configurational entropy. Using the LLCP location and the Kauzmann locus in the plotted phase diagram, we estimate $(T_c - T_K)/T_c \sim -0.031$. As a further example, a recent MEAM model of metallic gallium has been reported to exhibit a liquid–liquid critical point (LLCP) (49, 50). By digitizing the $P_{IS}(v)$ data from the GGT results, we obtain $(dP_{IS}/dv)_{max} = 0.4275$ (23) (Fig. S10d), indicating a pronounced first-order-like GGT signature. Using $T_c = 447.5\,K$ and $T_g = 350\,K$ for the low-density liquid, we obtain $(T_c - T_g)/T_c$ ~0.2179. This relatively large positive temperature separation is consistent with the LLCP remaining accessible above the glass-transition temperature and accompanies the pronounced GGT signature observed in this model. Because these literature estimates use different kinetic criteria and preparation protocols, the cross-model comparison is qualitative and does not establish a common quantitative calibration.

Among the systems examined, GGTs show a first-order-like transition when the LLCP is accessible before vitrification. These comparisons are consistent with the connection between LLCP accessibility and GGT sharpness found in FJ models. This relation can be understood from their difference in the schematic phase diagram (Fig.4). We propose that vitrification preserves not only structural features of the parent liquid, but also aspects of its mechanical stability. When $T_c > T_g$, the glass transition line crosses the liquid-liquid coexistence line. LLPT is accessible before vitrification, and the subsequent GGT crosses a mechanically unstable region, resulting in a subcritical, first-order-like transformation (Fig. 4a). When $T_c < T_g$, kinetic arrest occurs before the LLCP. Above the LLCP, the liquid can cross the Widom line (51) and evolve continuously from LDL-like to HDL-like behavior without a thermodynamic singularity (Fig. 4b). Such crossovers are often accompanied by maxima in thermodynamic response functions, such as the isothermal compressibility (29, 45). Similarly, the GGT connects LDA and HDA along a smooth pathway without an abrupt mechanical instability, resembling the continuous supercritical crossover of liquid (Fig. 4b). In this picture, LLCP accessibility is reflected not in whether the two amorphous states exist, but in the mechanical character of the pathway connecting them. Once a quantitative relationship has been established for real substances, the mechanical sharpness of the GGT could be used to estimate the LLCP temperature. For example, concentration-dependent GGT sharpness has been reported in hyperquenched LiCl–$H_2O$ solutions(52, 53), which provide a potential system in which to explore this approach.

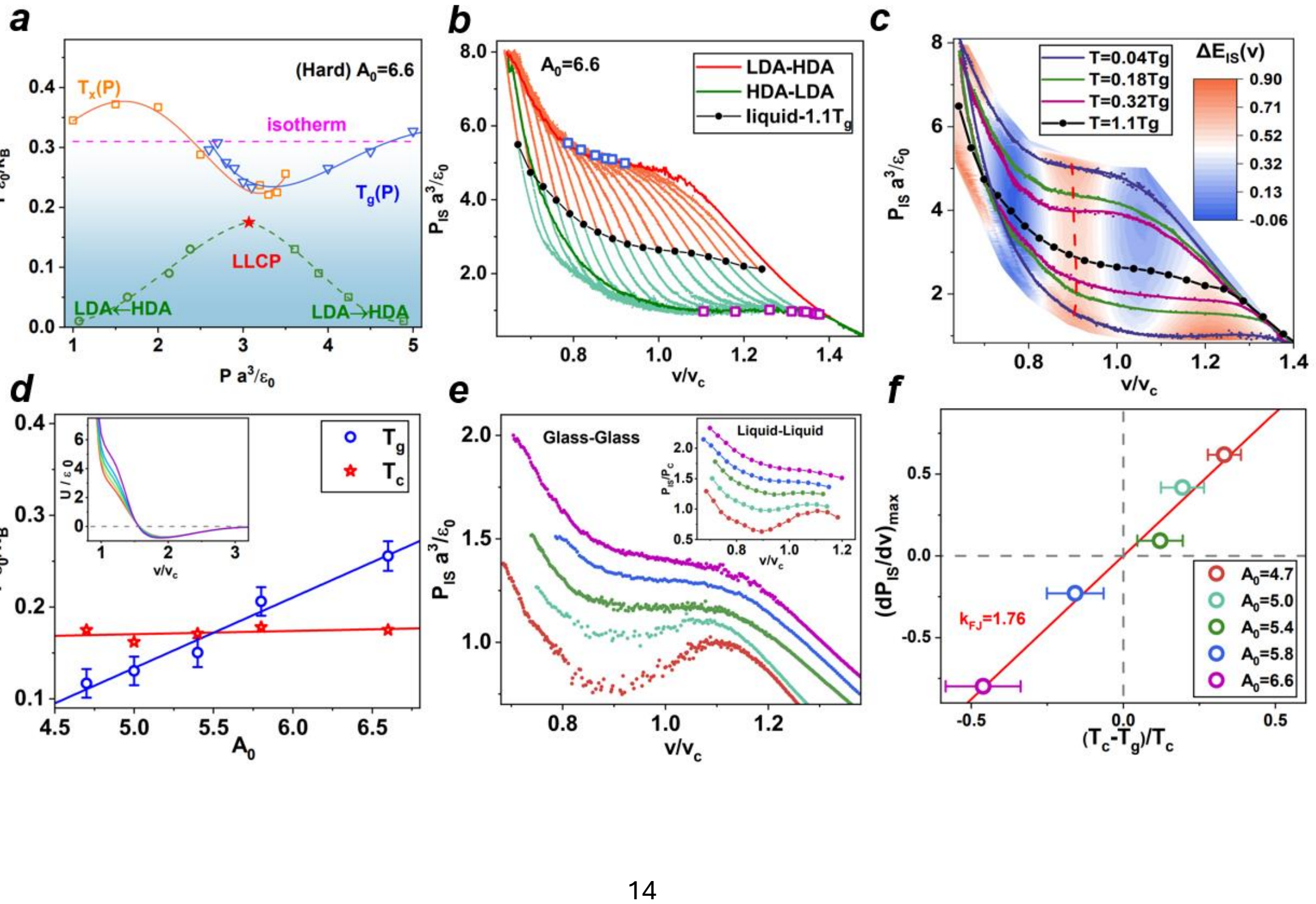

**Figure 3. LLCP stability and the mechanical character of glass–glass transitions. a,** Temperature–pressure (T–P) phase diagram of the $A_0$ = 6.6 FJ model, for which the LLCP lies below $T_\mathrm{g}$. The estimated LLCP ($T_c = 0.175$, $P_c = 3.07$, Methods), crystallization temperatures $T_\mathrm{x}$, glass-transition temperatures $T_\mathrm{g}$, and GGT states are indicated. **b,** $P_{IS}(v)$ for liquid-derived ISs compression and decompression trajectories at $A_0$ = 6.6. Blue and purple squares denote the corresponding transition points, and the conventional GGT pathway is shown for comparison. The transition points trace a smooth, monotonic GGT pathway. **c,** Relative inherent-structure energy $\Delta E_{IS}(v) = E_{IS}(v) - E_{IS,crystal}(v)$ landscape for $A_0$ = 6.6. Two distinct low- and high-density megabasins remain separated by a higher-energy ridge. GGT pathways at different temperatures below $T_\mathrm{g}$ are superimposed. d, Dependence of $T_\mathrm{g}$ and $T_\mathrm{c}$ on the interaction-softness parameter $A_0$, showing that continuous tuning of the same FJ interaction family shifts the LLCP from above to below the glass-transition temperature. Inset: pair potentials for different $A_0$. The same color scheme is used for the FJ models in panels d–f and the inset: red, light green, green, blue, and purple correspond to $A_0$=4.7, 5.0, 5.4, 5.8, and 6.6, respectively. e, Inherent-structure pressure $P_{IS}(v)$ along the low-temperature GGT at $T = 0.01\epsilon_0/k_B$ for $A_0$=4.7, 5.0, 5.4, 5.8, and 6.6, with curves vertically offset for clarity. The van der Waals–like loop progressively weakens and eventually disappears as the LLCP becomes kinetically inaccessible below $T_\mathrm{g}$. Inset: $P_{IS}(v)$ of the corresponding equilibrium liquids at $T = 1.1T_g$, showing a parallel evolution from a pronounced loop to a smooth, monotonic response. f, Maximum slope $(dP_{IS}(v)/dv)_{max}$ as a function of the LLCP-stability parameter $(T_c - T_g)/T_c$ for the FJ family. Positive values of $(T_c - T_g)/T_c$ indicate an LLCP accessible before vitrification, whereas negative values indicate an LLCP kinetically buried below $T_g$.

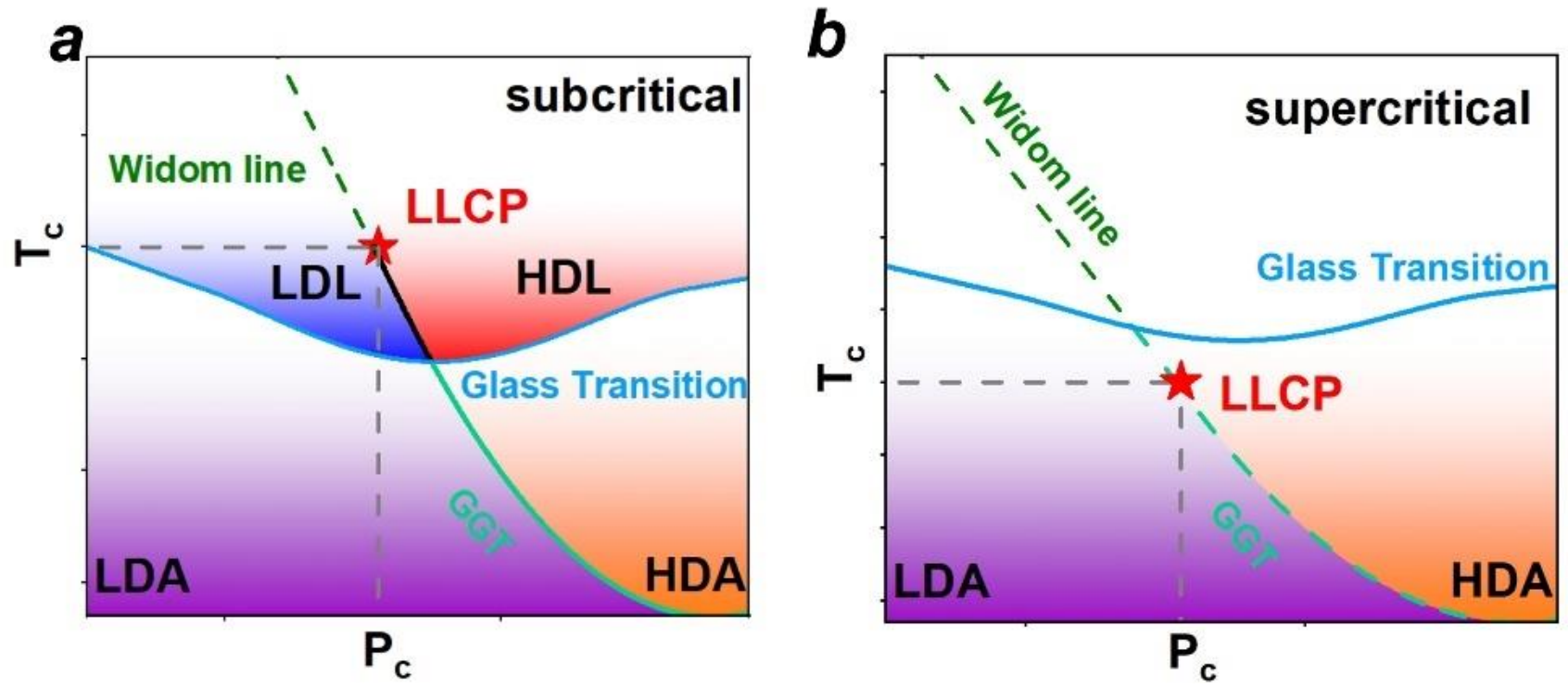

**Figure 4. Schematic phase diagrams illustrating two regimes of LLCP stability and GGT response.** a, Accessible LLCP ($T_\mathrm{g} < T_\mathrm{c}$), where the liquid–liquid transition remains accessible before vitrification. The subsequent transformation between low- and high-density amorphous states proceeds through a mechanically unstable pathway, giving rise to a first-order-like GGT. b, Kinetically buried LLCP ($T_\mathrm{g} > T_\mathrm{c}$), where vitrification occurs before the liquid reaches the critical region. The corresponding liquid regime exhibits a continuous supercritical crossover, represented schematically by the Widom line, while the subsequent GGT

proceeds along a smooth, mechanically stable pathway. The two regimes illustrate how the mechanical character of the GGT is associated with the stability of the parent liquid–liquid transition before vitrification.

## Summary and Discussion

In this work, we identify the mechanical origin of the glass–glass transition (GGT) in terms of the response of liquid-derived inherent structures (ISs) within the potential energy landscape. In a continuously tunable Fermi–Jagla model family, the transition points of independently prepared liquid-derived ISs trace the corresponding GGT pathways. This shows that the conventional GGT follows the stability limits of amorphous basins sampled by the parent liquid. The liquid- and glass-derived ISs also occupy corresponding low- and high-density megabasins, but the LLPT and GGT follow different pathways between them: the equilibrium liquid follows a lower-energy pathway, whereas the arrested glass follows a higher-energy nonequilibrium pathway because of kinetic arrest.

By varying the interaction softness, we further show that the mechanical character of the GGT changes systematically with LLCP stability. Here, LLCP stability refers to its position relative to the glass transition, rather than its thermodynamic stability relative to the crystalline phase. For softer interactions, $T_c > T_g$, so the LLCP remains accessible before vitrification and the subsequent GGT is first-order-like, with a mechanically unstable region along the IS pathway. For hard interactions, $T_c < T_g$, kinetic arrest occurs before the critical region is reached and the GGT becomes smooth without an abrupt mechanical instability. The low- and high-density megabasins remain present in both regimes, while the mechanical character of the connecting pathway changes systematically.

These results suggest that the mechanical response of a GGT may provide information about the liquid state before vitrification. Direct experimental observation of LLPTs is difficult because the deeply supercooled liquid is often inaccessible due to crystallization or vitrification. At low temperatures, thermal contributions to pressure are reduced, and the pressure–volume response may more closely reflect the underlying energy landscape. The predicted mechanical instability can be probed through bulk-modulus softening and hysteresis. Comparing these signatures with independent constraints on LLCP locations across glass-forming systems would provide a direct test of the proposed connection between GGT mechanics and liquid-state criticality.

## Acknowledgments

We thank Prof. Nicolas Giovambattista, Tanaka Hajime, Thomas Loerting, Francesco Sciortino for helpful discussions. Y. L. and L. X. acknowledge the National Natural Science Foundation of China under Grant No. 12535001, No. 11935002, No. 11525520; National Key R&D Program under grant 2021YFA1400500. G.S. acknowledges the National Natural Science Foundation of China under Grant No.12574217, No. 12535001. X.-L. Wang is grateful to support by the Research Grants Council of the Hong Kong Special Administrative Region, China (JLFS/P-101/24) and a grant at City University of Hong Kong (CityU Project No. 9229019). We are grateful for the computational resources provided by TianHe-1A and the High-Performance Computing Platform of Peking University.

## Data availability

The data that support the findings of this study are available from the corresponding author upon reasonable request.

## Methods

### Fermi-Jagla model

We employed the isotropic pairwise Fermi-Jagla (FJ) potential, a widely used model in studies of liquid-liquid phase transitions (LLPTs) and glass-glass transitions (GGTs)(44, 45). The FJ potential is modified from the original Jagla potential by replacing its linear ramp with a smooth Fermi function and the hard-sphere repulsion with a soft, power-law term. This model features two characteristic length scales: a steep repulsive core at $r = a$, and a soft-core repulsive shoulder for $a < r < 1.7a$, mimicking the core-softened interactions that underlie polymorphism in various systems.

The pairwise interaction is defined as:

$$U = \epsilon_0[(\frac{a}{r})^n + \frac{A_0}{1+\exp\left[\frac{A_0}{A_1}\left(\frac{r}{a}-A_2\right)\right]} - \frac{B_0}{1+\exp\left[\frac{B_0}{B_1}\left(\frac{r}{a}-B_2\right)\right]}],$$

where $n$, $A_i$, and $B_i$ ($i = 0,1,2$) are model parameters specified in Table 1. Here, $\epsilon_0$ and $a$ define the units of energy and length, respectively. The cutoff distance was $r_c = 4.0a$. Following Sun et al. (54), we added a switching correction to the Fermi–Jagla potential to ensure that the potential energy and force vanish smoothly at the cutoff. The switching correction was applied between $r = 0.9r_c$ and $r_c$, using a piecewise polynomial function.

To systematically investigate the relationship between LLPT and GGT behavior, we varied the interaction softness parameter $A_0$ over the range 4.7, 5.0, 5.4, 5.8, and 6.6. Increasing $A_0$ makes the interaction harder (Inset in Fig.3d). This tuning shifts the relative positions of the glass-transition temperature $T_g$ and the liquid–liquid critical temperature $T_c$ while preserving the same underlying interaction form.

| n | $A_0$ | $A_1$ | $A_2$ | $B_0$ | $B_1$ | $B_2$ |
|---|---|---|---|---|---|---|
| 20 | 4.7-6.6 | 0.72 | 1.36 | 1.00 | 0.28 | 2.36 |

**Table I**. Parameters of the pair interaction potential used in the Fermi-Jagla model. List of parameters for the repulsive and attractive components of the isotropic Fermi-Jagla pair potential as defined in the Methods.

**Molecular dynamics simulations**

Molecular dynamics (MD) simulations of the FJ model were performed using LAMMPS(55). Periodic boundary conditions were applied in all directions, and temperature and pressure were controlled using Nosé–Hoover thermostat and barostat. Simulations were performed for systems of N = 1728 particles in cubic boxes. For each interaction softness parameter $A_0$, the liquid–liquid critical point (LLCP), glass-transition temperatures $T_g(P)$, and crystallization temperatures $T_x(P)$ were determined from isochoric and isobaric cooling simulations. The slowest cooling rate used was $q_T = 10^{-5}\epsilon_0^{3/2}/(m^{1/2}k_B a)$. The integration time step was $\Delta t = 0.001$ in reduced units.

**Liquid–liquid critical point (LLCP)**

For the two representative models discussed in the main text, $A_0 = 4.7$ and $6.6$, equilibrium isochores were obtained from simulations performed at a series of fixed temperatures and specific volumes spanning $v = 1.0 - 4.0a^3$. At each state point, the system was equilibrated for $10^8$ MD steps, followed by another $10^8$ steps for sampling. The LLCP was identified from the convergence or crossing of neighboring isochores in the $P - T$ plane, together with the corresponding change in the $P(v)$ equation of state. For $A_0 = 4.7$, the LLCP is directly accessible in the equilibrium liquid and $T_c = 0.175\epsilon_0/k_B$ and $P_c = 0.528\epsilon_0/\mathrm{a}^3$ were determined from the highest-temperature crossing of neighboring isochores (Fig. S11a).

For $A_0 = 6.6$, the equilibrium isochores approach one another upon cooling but no common crossing is reached before kinetic arrest. We therefore estimate the LLCP location by extrapolating the accessible liquid isochores. Specifically, the equilibrium portions of neighboring isochores were fitted with third-order polynomials in $P(T)$ and extrapolated to lower temperature. A cubic polynomial was used as the lowest-order polynomial that adequately describes the curvature of the isochores. The highest-temperature convergence of the extrapolated isochores gives $T_c \sim 0.175\epsilon_0/k_B$ and $P_c \sim 3.07\epsilon_0/\mathrm{a}^3$ (Fig. S5).

For the intermediate models ($A_0 = 5.0 - 5.8$), we first performed slow isochoric cooling simulations to estimate the approximate locations of the LLCPs (Fig. S12). We then performed equilibrium MD simulations at selected temperatures and volumes near these regions to determine the liquid isochores more accurately (Fig. S5). The same criterion was applied: an accessible LLCP was identified from the convergence or crossing of neighboring equilibrium isochores, whereas a buried LLCP was estimated by extrapolating the equilibrium portions of the isochores below the onset of kinetic arrest.

We quantify the stability of the LLCP relative to vitrification by $(T_c - T_g)/T_c$, using $T_g = T_g(P_c)$ because the glass-transition temperature depends on pressure, with $P_c$ denoting the critical pressure of the LLCP. Positive values indicate that the LLCP is accessible before vitrification, whereas negative values indicate that it is kinetically buried below $T_g$. The determination of $T_g(P)$ and $T_x(P)$ is described below.

**Glass-transition and crystallization temperatures**

Isobaric cooling simulations were performed at a series of pressures, with the pressure range depending on the model, using the same cooling rate $q_T$. At low pressures, the system crystallizes upon cooling. The crystallization temperature $T_x(P)$ was identified from simultaneous abrupt changes in potential energy and density (or specific volume), together with the appearance of crystalline order. Operationally, $T_x$ was taken as the first sampled temperature at which the crystallized state was reached during cooling. At higher pressures, crystallization is suppressed and the system forms a glass. The enthalpy $H(T)$ was obtained during isobaric cooling, and the isobaric heat capacity was calculated as $C_P = \left(\frac{\partial H}{\partial T}\right)_P$. The glass-transition temperature $T_g(P)$ was taken as the temperature of the maximum in $C_P(T)$ (Fig. S1). For the LLCP-stability analysis, $T_g$ at $P_c$ was obtained directly from an isobaric cooling simulation at $P_c$ using the same criterion. The $H(T)$ and $C_P(T)$ curves were averaged over 10–20 independent cooling runs, depending on $A_0$. Although $T_g$ depends on cooling rate, the variation over the rates examined is small (Fig. S13-17), and values from the slowest isobaric cooling trajectories were used.

**Potential-energy-landscape analysis**

The potential energy landscape (PEL) describes the potential energy of a system of N particles as a function of their atomic coordinates in a 3N-dimensional configuration space. Different configurations correspond to different points on this landscape, and the system explores different regions of the PEL as it evolves. At high temperatures, thermal fluctuations allow the system to explore many basins, whereas at low temperatures the dynamics become increasingly confined to local basins.

Following the Stillinger–Weber potential-energy-landscape formalism(10, 38, 39), each basin is associated with a local minimum, termed an inherent structure (IS). For each configuration sampled along the LLPT and GGT pathways, we obtained the corresponding

IS by conjugate-gradient energy minimization. The energy and pressure of the minimized configuration are denoted $E_{IS}(v)$ and $P_{IS}(v)$, respectively, where $v$ is the specific volume.

**Liquid-state sampling**

To compare liquid-derived and glass-derived inherent structures, equilibrium liquid configurations were sampled at $T = 1.1T_g$ over the specific-volume range v = 1.0-4.0a³. Starting from high-temperature configurations at $T = 0.6\epsilon_0/k_B$, the system was equilibrated for $10^8$ steps, followed by an additional $10^8$ steps for data collection. The liquid-state behavior was examined using the equilibrium equation of state $P(v)$. The LLPT was identified from nonmonotonic behavior in the equilibrium equation of state $P(v)$. A van der Waals–like loop in $P(v)$ and $P_{IS}(v)$ is consistent with a transition between low- and high-density liquid states.

**Glass–glass transition preparation and transition pressure**

To characterize the glass–glass transition (GGT), we prepared LDA by decompressing HDA rather than using the LDA obtained directly from the initial cooling procedure, whose structure and energy depend sensitively on the cooling rate. First, ten independent LDL configurations were cooled to $T = 0.01\epsilon_0/k_B$ at a rate $q_T = 10^{-3}\epsilon_0^{3/2}/(m^{1/2}k_B a)$ to prepare LDA. The LDA states were then compressed to form HDA. GGT trajectories reported in the main text were obtained from inherent structures (ISs) sampled during a second decompression–compression cycle. The compression and decompression rate was $q_P = 10^{-3}\epsilon_0^{\frac{3}{2}}/(m^{\frac{1}{2}}a^4)$, and ISs were sampled at pressure intervals $\Delta P = 0.02 - 0.08\ \epsilon_0/a^3$, depending on the pressure range. The reported $P_{IS}(v)$ is averaged over ten independent runs. At each temperature, we define the GGT pressures $P_{LDA-HDA}(T)$ and $P_{HDA-LDA}(T)$ as the pressures defined by the maximum slope of the mean squared displacement (Figs. 1a and 3a). For trajectories initiated from liquid-derived ISs, transition points were identified when the MSD reached a fixed threshold of $0.25\,a^2$ during compression and $1.0a^2$ during decompression. The corresponding pressure was taken as the transition pressure. These threshold points lie near the region of the steepest MSD increase and rapid volume change.

**Liquid-derived inherent structures and compression–decompression simulations**

To characterize how the response of liquid-derived ISs depends on their initial volume, equilibrium liquid configurations were sampled at a series of specific volumes $v$ at $T = 1.1T_g$. At each volume, ten temporally separated equilibrium configurations were directly energy-minimized to obtain the corresponding liquid-derived ISs. The minimized configurations were then assigned velocities corresponding to $T = 0.01\epsilon_0/k_B$ before the subsequent compression–decompression simulations. For each initial volume, ten independent compression–decompression runs were performed using these liquid-derived ISs as initial configurations. The runs showed similar behavior, and the reported $P_{IS}(v)$ and $E_{IS}(v)$ values are averages over the ten runs.

**Structural and mechanical observables**

Structural rearrangements during compression and decompression were tracked using the mean-square displacement (MSD; Fig. S18) relative to the initial configuration of each run.

$$MSD(t) = \frac{1}{N}\sum_i |\boldsymbol{r}_i(t) - \boldsymbol{r}_i(0)|^2$$

Because the simulation volume changes along these paths, the MSD includes both affine displacements due to changes in the box size and non-affine particle motion. A separate check showed that the increase near the GGT persists after removing the affine contribution. Particle displacements were calculated with periodic boundary conditions, with the system center kept fixed. The bulk modulus was calculated from the finite-temperature pressure response as $B = -v(\frac{\partial P}{\partial v})_T$ (Fig. S19). For each set of ten independent runs, $P(v)$ was first averaged and then fitted with a smooth polynomial before taking the derivative. The minimum in $B$ marks the strongest mechanical softening along the transformation. After the transition, the self-intermediate scattering function (Fig. S20) decays to approximately $1/e^2$, indicating substantial structural decorrelation from the initial configuration. The wavevector ($k_0$) used in calculation of intermediate scattering function was estimated as ($2\pi/r_{peak}$), where $r_{peak}$ is the position of the first peak in radial distribution function, providing an estimate of the wavevector corresponding to the principal peak of S(q).

**Energetic reference states**

To quantify the energetic response along compression and decompression pathways, we define $\Delta E(v) = E_{\mathrm{IS}}(v) - E_{\mathrm{IS,liquid}}(v)$ . At each specific volume, 100 independent equilibrium liquid configurations were sampled at $T = 1.1T_g$ and energy-minimized to obtain their IS energies. The mean IS energy was then calculated for each volume, and $E_{\mathrm{IS,liquid}}(v)$ was obtained by fitting these mean values as a function of (v). The $E_{\mathrm{IS,liquid}}(v)$ reference curve was used for both compression and decompression. For small deformations around an initial liquid-derived IS configuration, $\Delta E(v)$ is approximately quadratic in the volume change. Deviations from this local quadratic response are analyzed together with the displacement and pressure responses.

For a separate comparison of amorphous and crystalline landscape energies, crystalline reference states were prepared at a sequence of fixed specific volumes using slow cooling and low-temperature compression or decompression. At each volume, the lowest-energy crystalline configuration obtained at $T = 0.01\epsilon_0/k_B$ was energy-minimized to define $E_{IS,crystal}(v)$ (Fig. S4). The excess energy $\Delta E_{\mathrm{IS}}(v)$=$E_{\mathrm{IS}}(v) - E_{IS,crystal}(v)$ quantifies the energetic separation of the amorphous states from the crystalline reference and enables a consistent comparison between liquid-derived and glass-derived inherent structures.

## Reference

1. M. Henkel, *Non-equilibrium phase transitions* (Springer, 2008).
2. P. W. Anderson, More is different: Broken symmetry and the nature of the hierarchical structure of science. *Science* **177**, 393-396 (1972).
3. G. Parisi, The order parameter for spin glasses: a function on the interval 0-1. *Journal of Physics A: Mathematical and General* **13**, 1101 (1980).
4. S. Zhang *et al.*, Polymorphism in glassy silicon: Inherited from liquid-liquid phase transition in supercooled liquid. *Scientific Reports* **5**, 8590 (2015).
5. L. Li *et al.*, Liquid-liquid transition and inherited signatures in Zr-Cu-Ni-Al metallic glasses. *Journal of Non-Crystalline Solids* **600**, 122029 (2023).
6. T. E. Gartner III, S. Torquato, R. Car, P. G. Debenedetti, Manifestations of metastable criticality in the long-range structure of model water glasses. *Nature Communications* **12** (2021).
7. S. Mossa, E. La Nave, F. Sciortino, P. Tartaglia, Aging and energy landscapes: application to liquids and glasses. *European Physical Journal B* **30**, 351-355 (2002).
8. S. Lan *et al.*, Hidden amorphous phase and reentrant supercooled liquid in Pd-Ni-P metallic glasses. *Nature Communications* **8** (2017).
9. Y. Jin, P. Urbani, F. Zamponi, H. Yoshino, A stability-reversibility map unifies elasticity, plasticity, yielding, and jamming in hard sphere glasses. *Science Advances* **4**, eaat6387 (2018).
10. P. G. Debenedetti, F. H. Stillinger, Supercooled liquids and the glass transition. *Nature* **410**, 259-267 (2001).
11. C. A. Angell, Relaxation in liquids, polymers and plastic crystals - strong fragile patterns and problems. *Journal of Non-Crystalline Solids* **131**, 13-31 (1991).

12. D. Ma *et al.*, Elastic moduli inheritance and the weakest link in bulk metallic glasses. *Physical Review Letters* **108**, 085501 (2012).
13. T. Loerting, V. V. Brazhkin, T. Morishita, Multiple amorphous-amorphous transitions. *Advances in Chemical Physics* **143**, 29-82 (2009).
14. O. Mishima, L. D. Calvert, E. Whalley, An apparently first-order transition between two amorphous phases of ice induced by pressure. *Nature* **314**, 76-78 (1985).
15. O. Mishima, Reversible first-order transition between two H2O amorphs at ~0.2 GPa and ~135 K. *Journal of Chemical Physics* **100**, 5910-5912 (1994).
16. O. Mishima, Relationship between melting and amorphization of ice. *Nature* **384**, 546-549 (1996).
17. O. Mishima, Polyamorphism in water. *Proceedings of the Japan Academy Series B* **86**, 165-175 (2010).
18. C. Tien, E. V. Charnaya, W. Y. Wang, Y. A. Kumzerov, D. Michel, Possible liquid-liquid transition of gallium confined in opal. *Physical Review B* **74** (2006).
19. Q. S. Zeng *et al.*, Properties of polyamorphous Ce-Al metallic glasses. *Physical Review B* **82** (2010).
20. Q. S. Zeng *et al.*, Origin of pressure-induced polyamorphism in Ce-Al metallic glass. *Physical Review Letters* **104** (2010).
21. M. C. Wilding, P. F. McMillan, Polyamorphic transitions in yttria-alumina liquids. *Journal of Non-Crystalline Solids* **293**, 357-365 (2001).
22. S. Lan *et al.*, Structural crossover in a supercooled metallic liquid and the link to a liquid-to-liquid phase transition. *Appl Phys Lett* **108** (2016).
23. Y. Z. Liu, G. Sun, L. M. Xu, Glass polyamorphism in gallium: Two amorphous solid states and their transformation on the potential energy landscape. *Journal of Chemical Physics* **154** (2021).
24. P. H. Poole, F. Sciortino, U. Essmann, H. E. Stanley, Phase behavior of metastable water. *Nature* **360**, 324-328 (1992).
25. P. G. Debenedetti, F. Sciortino, G. H. Zerze, Second critical point in two realistic models of water. *Science* **369**, 289–292 (2020).
26. J. C. Palmer, P. H. Poole, F. Sciortino, P. G. Debenedetti, Advances in computational studies of the liquid-liquid transition in water and water-like models. *Chemical Reviews* **118**, 9129-9151 (2018).
27. P. Gallo *et al.*, Water: A tale of two liquids. *Chemical Reviews* **116**, 7463-7500 (2016).
28. H. Tanaka, Liquid-liquid transition and polyamorphism. *Journal of Chemical Physics* **153** (2020).
29. Y. Z. Liu *et al.*, Nuclear quantum effects on the thermodynamic response functions of a polymorphic waterlike monatomic liquid. *Physical Review Research* **2** (2020).
30. G. N. Greaves *et al.*, Detection of first-order liquid/liquid phase transitions in yttrium oxide-aluminum oxide melts. *Science* **322**, 566-570 (2008).
31. S. Aasland, P. F. McMillan, Density-Driven Liquid-Liquid Phase-Separation in the System Al2o3-Y2o3. *Nature* **369**, 633-636 (1994).
32. N. Giovambattista, T. Loerting, B. R. Lukanov, F. W. Starr, Interplay of the Glass Transition and the Liquid-Liquid Phase Transition in Water. *Scientific Reports* **2** (2012).

33. A. Gordon, N. Giovambattista, Glass Transitions in a Monatomic Liquid with Two Glassy States. *Physical Review Letters* **112** (2014).
34. K. H. Kim *et al.*, Experimental observation of the liquid-liquid transition in bulk supercooled water under pressure. *Science* **370**, 978 (2020).
35. K. Amann-Winkel *et al.*, Liquid-liquid phase separation in supercooled water from ultrafast heating of low-density amorphous ice. *Nature Communications* **14** (2023).
36. M. Yang, T. Karmakar, M. Parrinello, Liquid-liquid critical point in phosphorus. *Physical Review Letters* **127**, 080603 (2021).
37. L. Henry *et al.*, Liquid-liquid transition and critical point in sulfur. *Nature* **584**, 382 (2020).
38. F. Sciortino, Potential energy landscape description of supercooled liquids and glasses. *Journal of Statistical Mechanics-Theory and Experiment* (2005).
39. F. H. Stillinger, A topographic view of supercooled liquids and glass formation. *Science* **267**, 1935-1939 (1995).
40. W. Kob, F. Sciortino, P. Tartaglia, Aging as dynamics in configuration space. *Europhys Lett* **49**, 590-596 (2000).
41. N. Giovambattista, H. E. Stanley, F. Sciortino, Potential-energy landscape study of the amorphous-amorphous transformation in H2O. *Physical Review Letters* **91** (2003).
42. N. Giovambattista, F. Sciortino, F. W. Starr, P. H. Poole, Potential energy landscape of the apparent first-order phase transition between low-density and high-density amorphous ice. *Journal of Chemical Physics* **145** (2016).
43. P. H. Handle, F. Sciortino, N. Giovambattista, Glass polymorphism in TIP4P/2005 water: A description based on the potential energy landscape formalism. *Journal of Chemical Physics* **150** (2019).
44. G. Sun, L. M. Xu, N. Giovambattista, Anomalous Features in the Potential Energy Landscape of a Waterlike Monatomic Model with Liquid and Glass Polymorphism. *Physical Review Letters* **120** (2018).
45. J. Y. Abraham, S. V. Buldyrev, N. Giovambattista, Liquid and glass polymorphism in a monatomic system with isotropic, smooth pair interactions. *Journal of Physical Chemistry B* **115**, 14229-14239 (2011).
46. T. A. Kesselring, G. Franzese, S. V. Buldyrev, H. J. Herrmann, H. E. Stanley, Nanoscale Dynamics of Phase Flipping in Water near its Hypothesized Liquid-Liquid Critical Point. *Scientific Reports* **2** (2012).
47. F. Sciortino, E. La Nave, P. Tartaglia, Physics of the liquid-liquid critical point. *Physical Review Letters* **91**, 155701 (2003).
48. N. Giovambattista, C. A. Angell, F. Sciortino, H. E. Stanley, Structural relaxation in the glass transition region of water. *Physical Review E* **72** (2005).
49. D. A. C. Jara, M. F. Michelon, A. Antonelli, M. de Koning, Theoretical evidence for a first-order liquid-liquid phase transition in gallium. *Journal of Chemical Physics* **130** (2009).
50. R. Z. Li, G. Sun, L. M. Xu, Anomalous properties and the liquid-liquid phase transition in gallium. *Journal of Chemical Physics* **145** (2016).
51. L. M. Xu *et al.*, Relation between the Widom line and the dynamic crossover in systems with a liquid-liquid phase transition. *Proceedings of the National Academy of Sciences* **102**, 16558-16562 (2005).

52. J. Giebelmann, J. Bachler, T. Loerting, Glass Polymorphism in Hyperquenched Aqueous LiCl Solutions. *Journal of Physical Chemistry B* **127**, 3463-3477 (2023).
53. J. Giebelmann, J. Bachler, T. Loerting, Isocompositional liquid-liquid transition in dilute aqueous LiCl solutions. *Physical Review Research* **6**, 023108 (2024).
54. G. Sun, L. M. Xu, N. Giovambattista, Relationship between the potential energy landscape and the dynamic crossover in a water-like monatomic liquid with a liquid-liquid phase transition. *Journal of Chemical Physics* **146** (2017).
55. S. Plimpton, Fast parallel algorithms for short-range molecular-dynamics. *Journal of Computational Physics* **117**, 1-19 (1995).

# Supplementary Information of: Mechanical Signatures of Liquid Polymorphism in Glass–Glass Transitions

*Yizhi Liu[1], Yixiang Chen[1], Shaoqin Jiang[1], Xun-Li Wang[2*], Gang Sun[3*],and Limei Xu[1,4,5*]*

*1 International Center for Quantum Materials, School of Physics, Peking University, Beijing 100871, China*

*2 Department of Physics and Center for Neutron Scattering, CityUHK, 83 Tat Chee Avenue, Kowloon, Hong Kong, China*

*3 Center for Advanced Quantum Studies, School of Physics and Astronomy, Beijing Normal University, Beijing 100875, China*

*4 Interdisciplinary Institute of Light-Element Quantum Materials and Research Center, Peking University, Beijing 100871, China*

*5 Collaborative Innovation Center of Quantum Matter, Beijing, China*

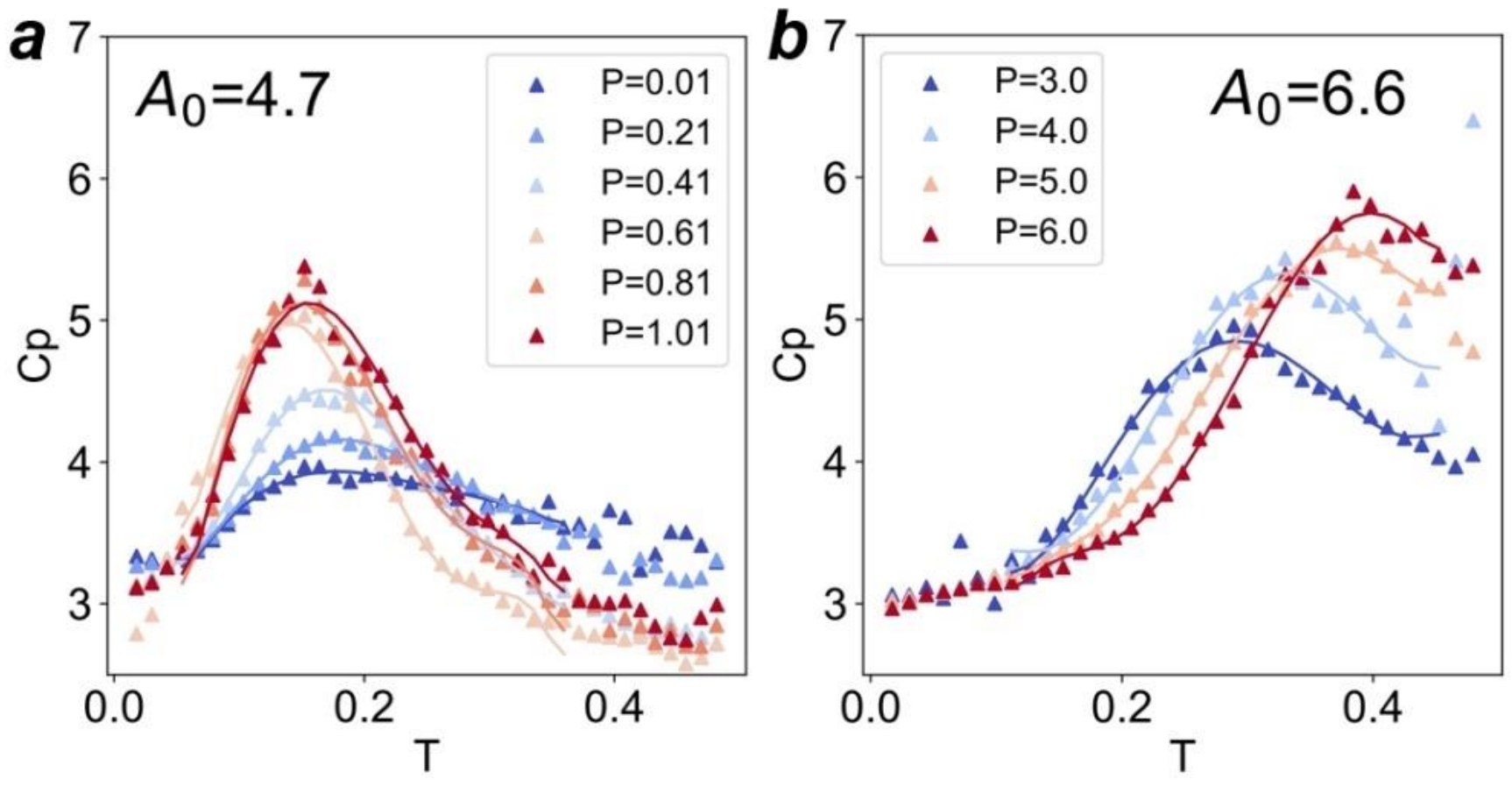

**Figure. S1. Heat capacity Cp during isobaric cooling** (cooling rate $q_T = 10^{-3}$ $\epsilon_0^{\frac{3}{2}}/m^{\frac{1}{2}}k_B a$) for the FJ liquid at different pressures (symbols): (a) $A_0$=4.7 and (b) $A_0$=6.6. The temperature corresponding to the maximum of $C_p$ shifts with pressure. Around each maximum, the data are fitted (solid lines) to determine the glass transition temperature $T_g$.

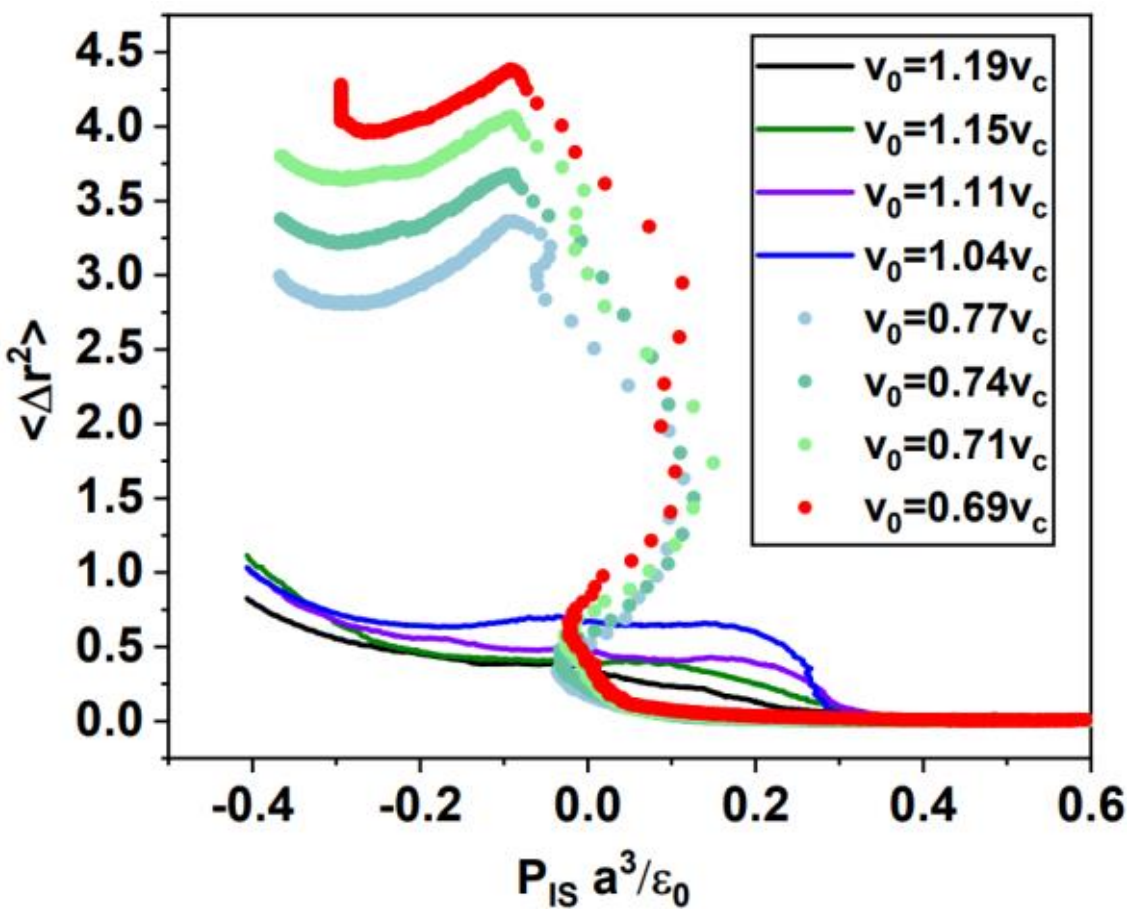

**Figure. S2. The MSD ($A_0 = 4.7$) along the decompression paths.** Circle-marked curves denote paths initiated from HDL-derived IS states, whereas solid curves denote paths initiated from LDL-derived IS states.

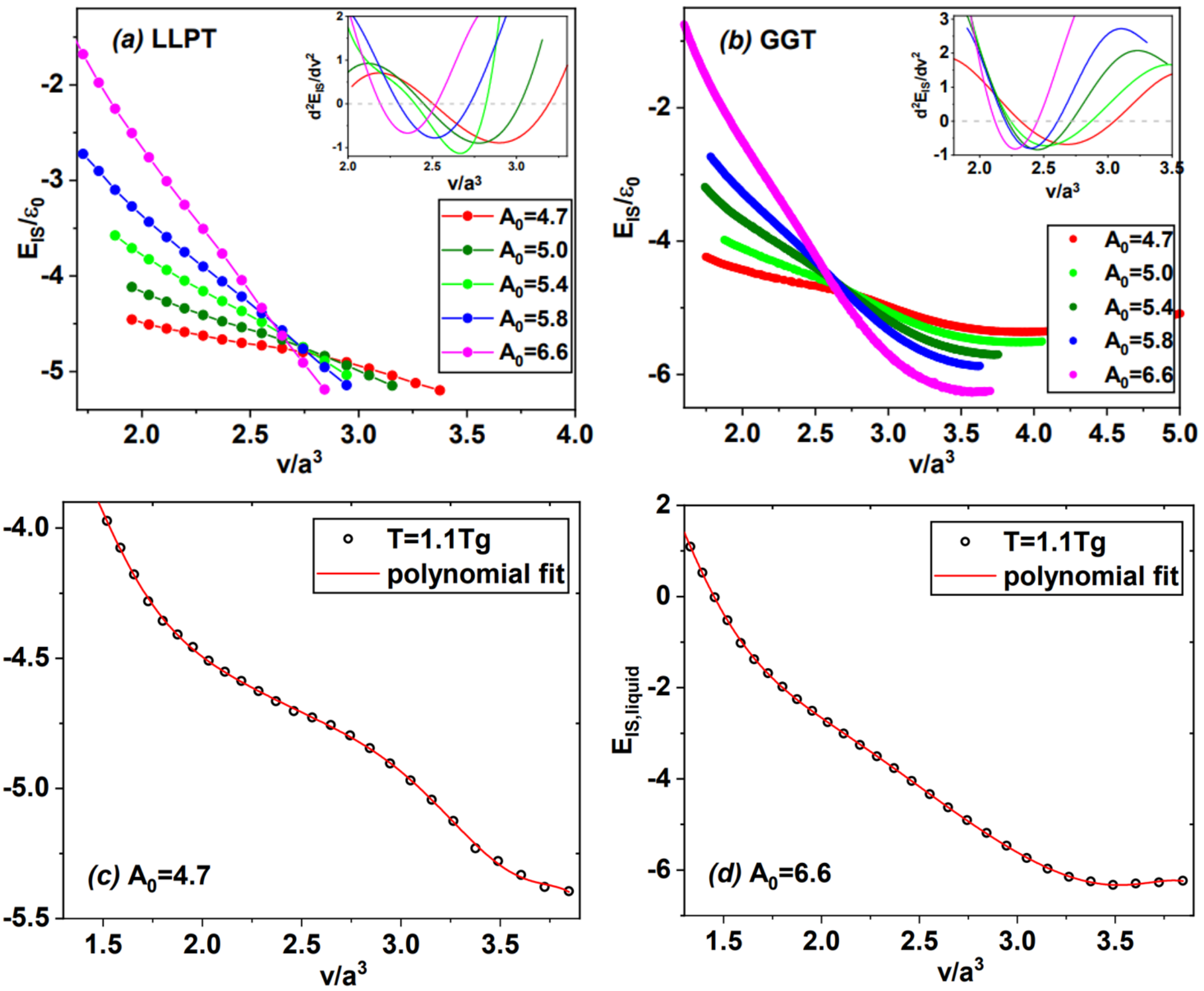

**Figure. S3. Potential energy of inherent structures**, $E_{IS}$(v), as a function of volume during (a) Liquid-state sampling at $T = 1.1T_g$ and (b) glass-glass transformation at $T = 0.01\epsilon_0/k_B$. Insets in both panels show the second derivative $(\partial^2\mathrm{E}/\partial \mathrm{v}^2)_T$ during the transitions. For all FJ systems ($A_0$= 4.7 − 6.6), $E_{IS}(v)$ exhibits a negative curvature. Negative curvature in $E_{IS}$ is observed for both sharp and smooth GGTs and is therefore not a sufficient condition for a first-order-like GGT. (c,d) Potential energy $E_{IS,liquid}(v)$ of inherent structures sampled by LLPT for $A_0$= 4.7 and 6.6 are shown as black circles, and the red lines represent fitted curves.

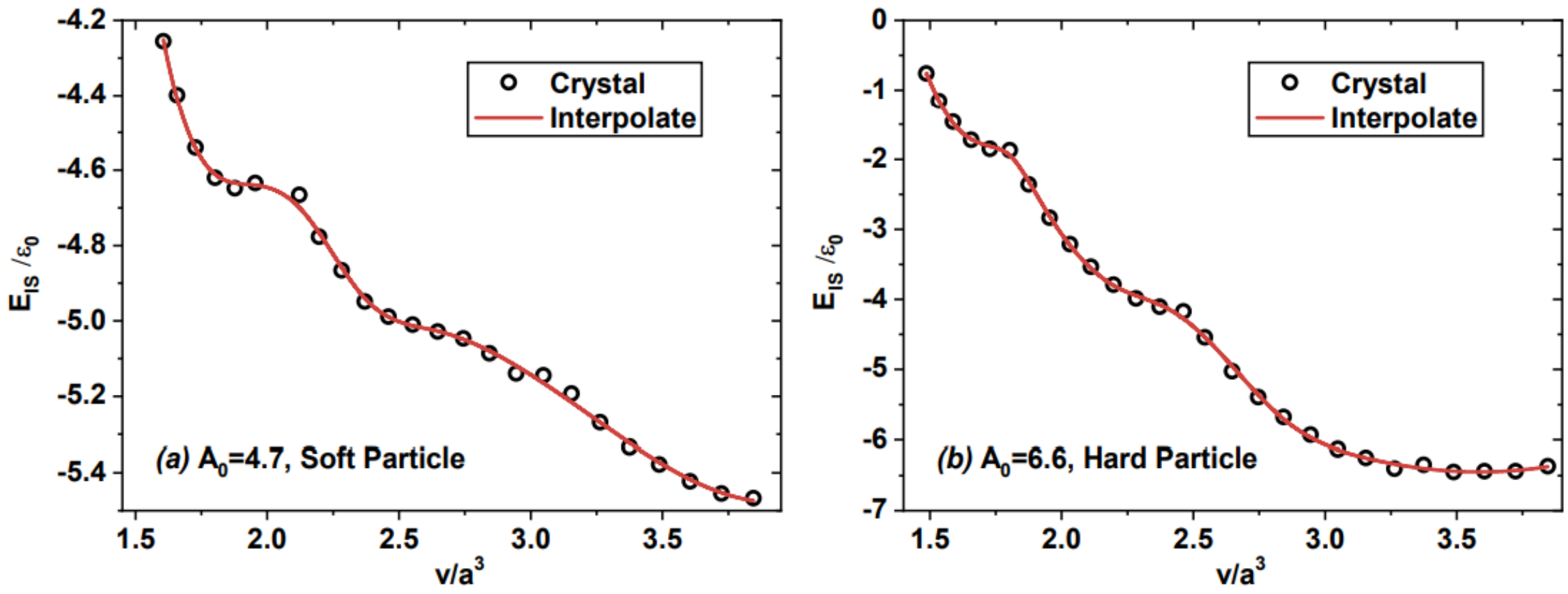

**Figure. S4. Potential energy $E_{IS}(v)$ of inherent structures sampled by crystals at different volumes.** Simulation data are shown as black circles, and the red lines represent interpolated results for comparison.

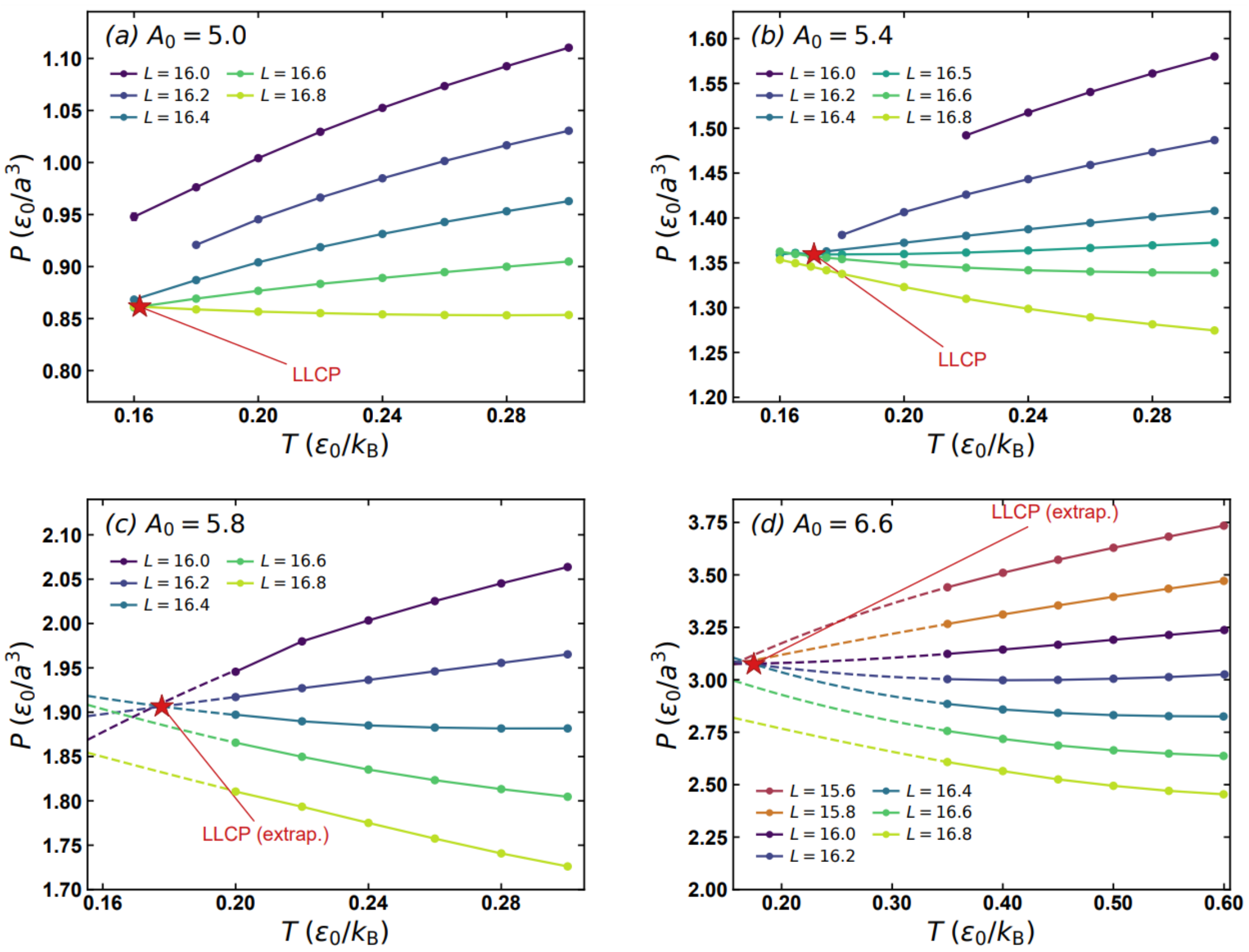

**Figure S5. Pressure–temperature isochores and estimated liquid–liquid critical points.** Panels (a–d) correspond to $A_0$=5.0, 5.4, 5.8, and 6.6, respectively. Circles denote retained pressure averages at the box lengths (L) indicated in each panel. Crystalline points are excluded. Red stars mark the LLCP. For $L = 16.4$ in (b), the $T = 0.170$ and $0.175$ means are obtained from stationary early precrystallization plateaus (metastable liquid). In (a,b), their locations are determined by linear interpolation between adjacent retained means, selecting the first crossing encountered upon cooling. In (c,d), dashed lines show cubic fits extrapolated from all retained points over $T = 0.20 - 0.30$ and $T = 0.35 - 0.60$, respectively. The resulting $(T_c, P_c)$ estimates are $(0.16182, 0.86141), (0.17100, 1.35926), (0.17778, 1.90642)$, and $(0.17496, 3.07393)$ for (a–d), respectively. The units of temperature and pressure are $\epsilon_0/k_B$ and $\epsilon_0/a^3$.

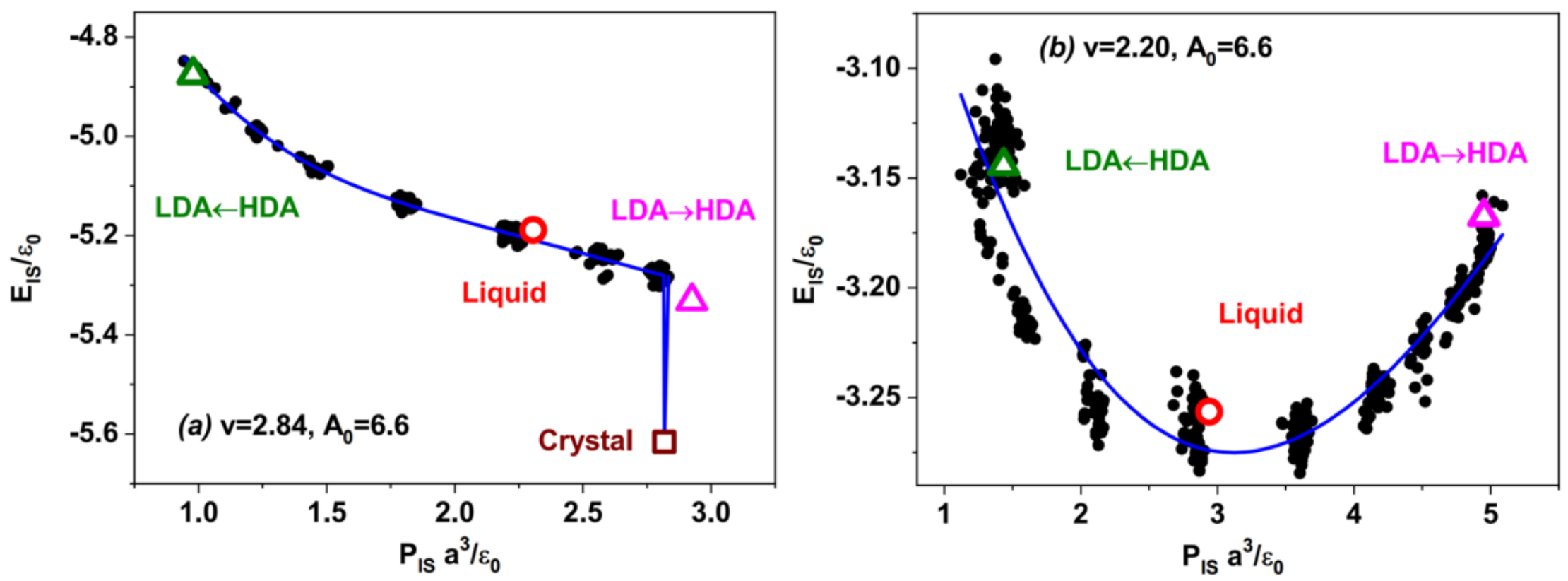

**Figure. S6. Energy of inherent structures**, $E_{\mathrm{IS}}(P_{\mathrm{IS}})$, sampled at (a) low density ($v = 2.84a^3$) and (b) high density ($v = 2.20a^3$) in hard particle system ($A_0 = 6.6$). ISs sampled from the liquid (red circle), LDA-HDA transformation (green triangle), HDA-LDA transformation (violet triangle), and crystal (wine square).

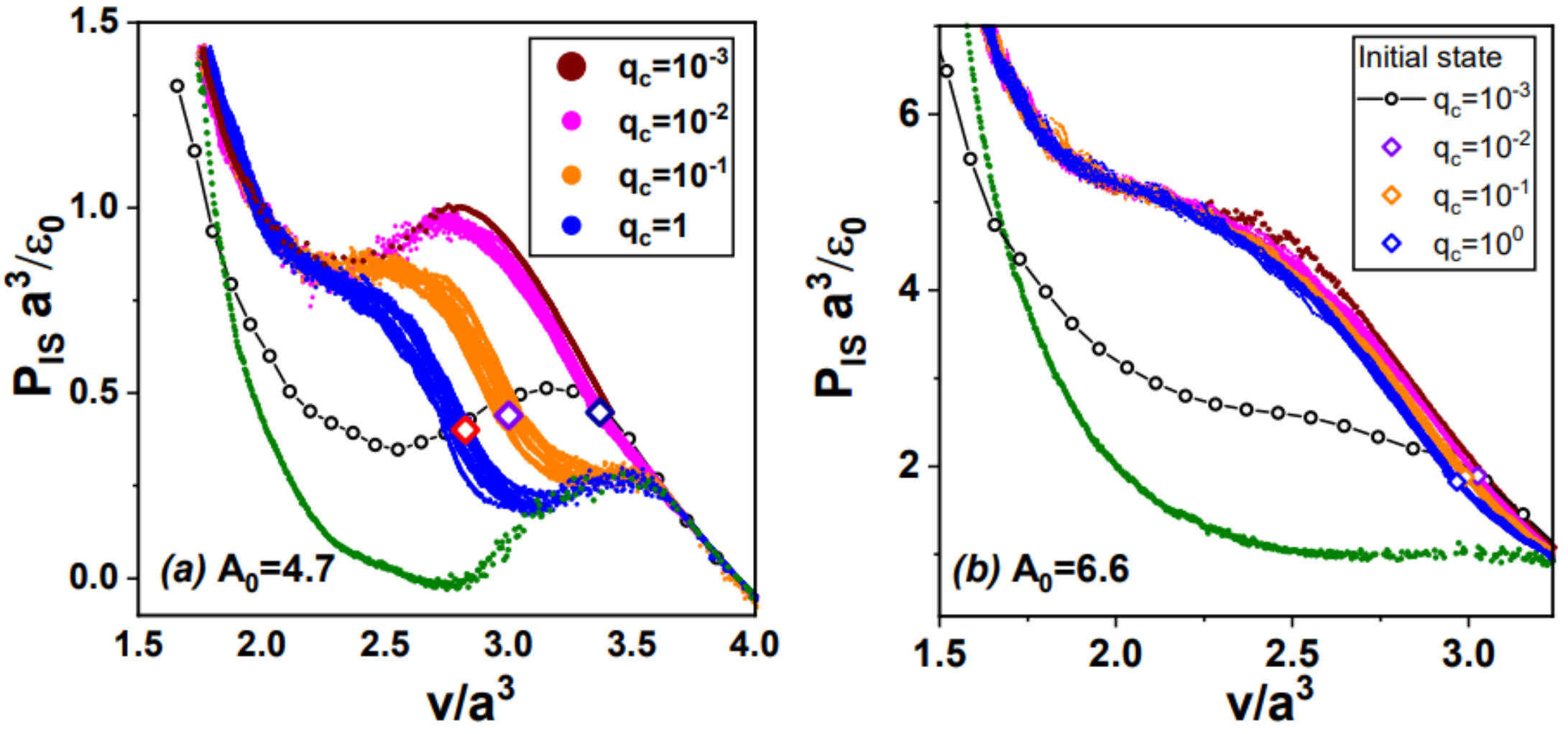

**Figure. S7. Pressure $P_{\mathrm{IS}}(v)$ of the inherent structures** sampled during compressing and decompressing LDAs prepared with different cooling rates for (a) $A_0$ = 4.7 and (b) 6.6. In each cooling rate, we show 10 independent simulations. This suggests that glasses obtained by rapid cooling can be regarded as a frozen liquid at different volumes.

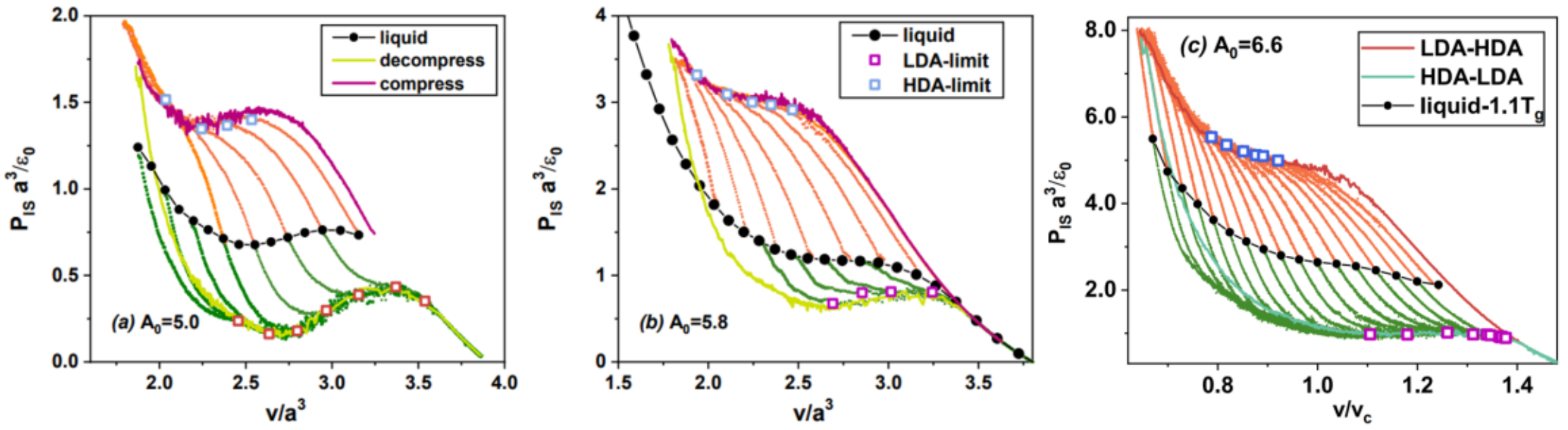

**Figure. S8. Pressure $P_{IS}(v)$ of the inherent structures** sampled during pressure-induced transformation starting from different liquid inherent structures in the FJ model with (a) $A_0$ = 5.0, (b) $A_0$ = 5.8 and (c) $A_0$ = 6.6. Open squares mark transition points, defined when the MSD reached a fixed threshold of $0.25a^2$ during compression and $1.0a^2$ during decompression. These threshold points lie near the region of the steepest MSD increase and rapid volume change.

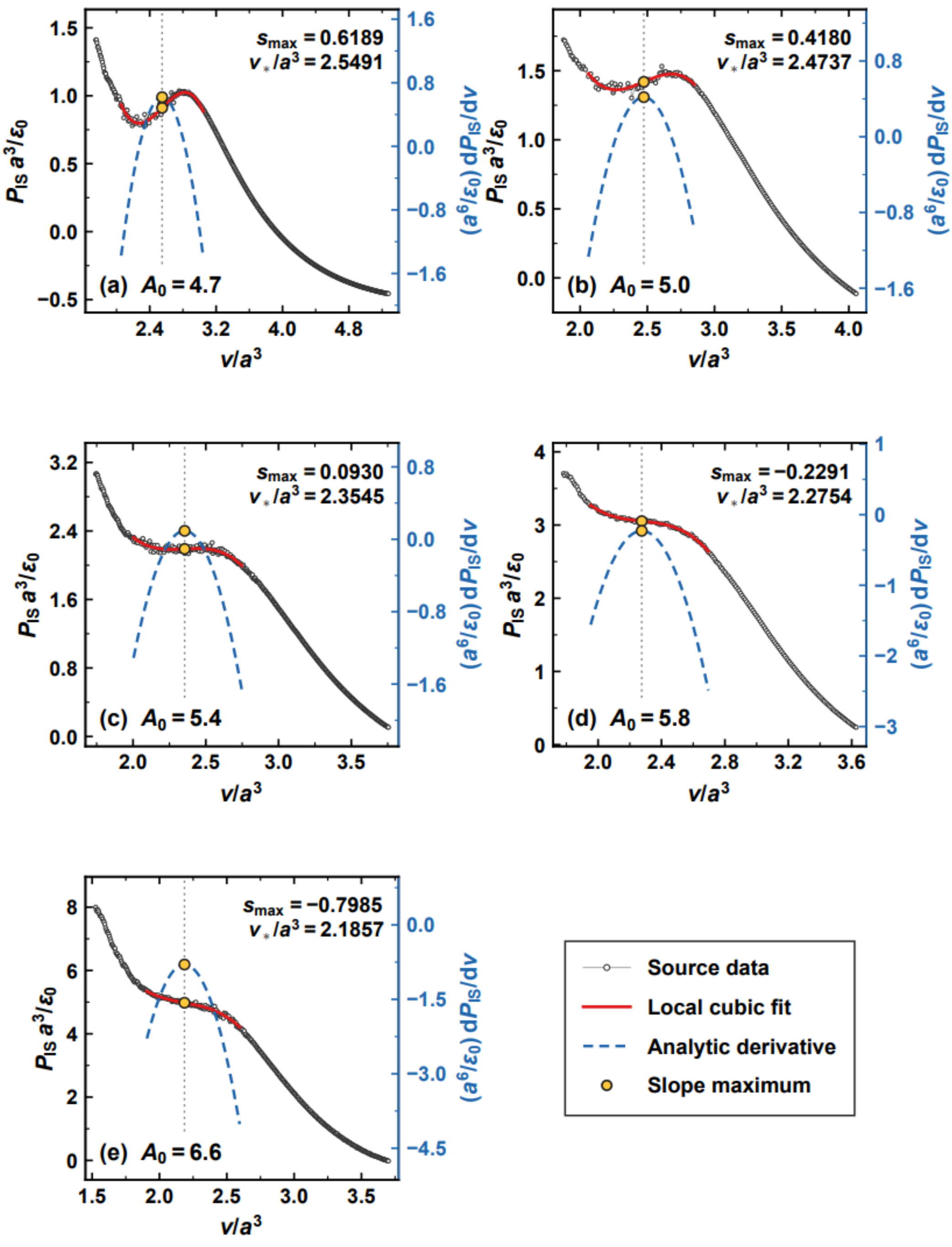

**Figure S9. Inherent-structure pressure–volume curves and local slope maxima for different interaction potentials.** Panels (a)–(e) correspond to $A_0 = 4.7, 5.0, 5.4, 5.8$ and 6.6, respectively. Open circles show the simulation data at $T = 0.01\epsilon_0/k_B$, red curves show cubic fits within the selected local intervals. Blue dashed segments show the analytic derivatives $dP_{IS}/dv$. Yellow circles mark the fitted maximum slope $s_{max}$ at $v_*$.

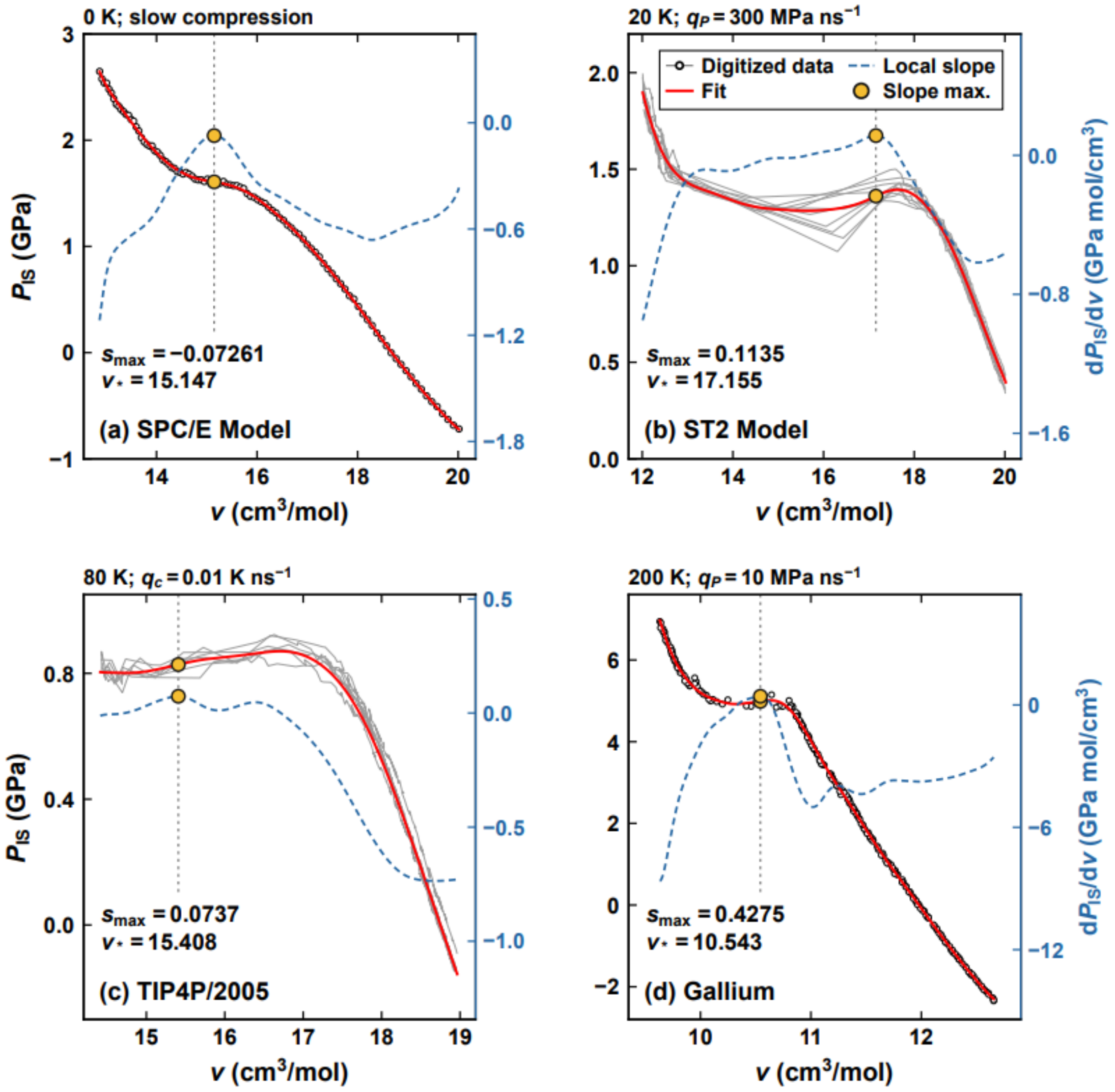

**Figure. S10. Inherent-structure pressure and local slopes during compression in four model systems.** (a) SPC/E water at 0 K, using the slow-compression branch in Fig. 2 of (1). (b) ST2 water at 20 K and a pressure-increase rate of 300 MPa ns$^{-1}$, using the compression trajectories in Fig. 1a of (2). (c) TIP4P/2005 water at 80 K, prepared at a cooling rate of 0.01 K/ns and compressed at 10 MPa/ns, using the compression trajectories in Fig. 10a of (3). (d) Gallium at 200 K and 10 MPa ns$^{-1}$, using the compression trajectory in Fig. 2(a) of (4). Black/gray lines and open circles show digitized published curves or markers. Red curves show polynomial pressure fits. Blue dashed segments show the analytic derivatives $dP_{IS}/dv$. Yellow circles mark the fitted maximum slope $s_{max}$ at $v_*$ volume. The units of $s_{max}$ and $v$ are given in GPa mol/cm$^3$ and cm$^3$/mol, respectively.

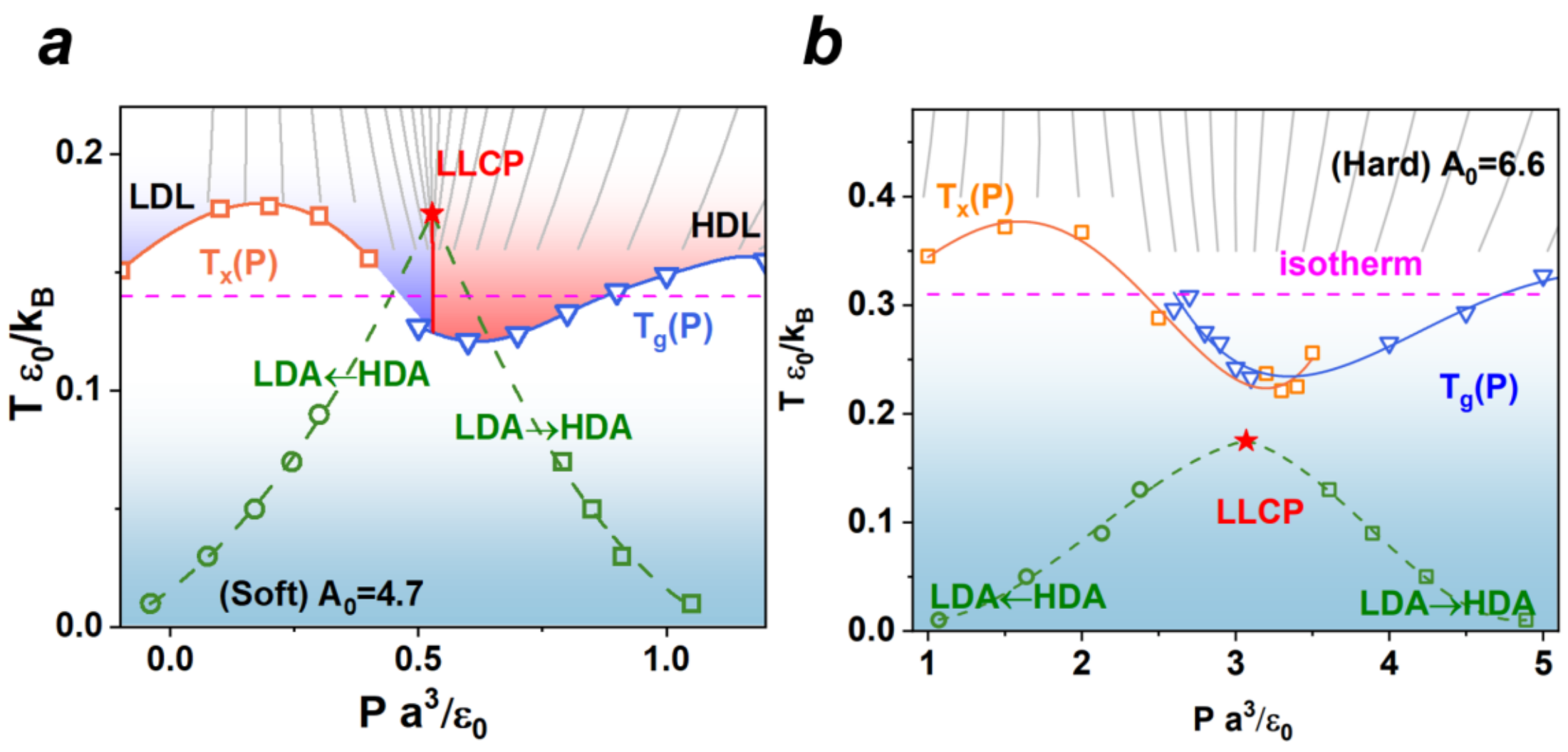

**Figure. S11. T–P phase diagrams for the FJ liquid** with (a) $A_0$ = 4.7 and (b) 6.6. When $A_0$ = 4.7, a liquid–liquid critical point (LLCP) is located at $T_C$ = 0.175 $\epsilon_0/k_B$, whereas no LLCP is observed for $A_0$ = 6.6. In the latter case, the LLCP position is estimated by extrapolating isochores. Orange squares denote crystallization points; blue triangles indicate the glass transition temperature ($T_g$); green circles and squares mark states associated with glass–glass transitions (GGTs). Gray lines represent isochores for different volumes.

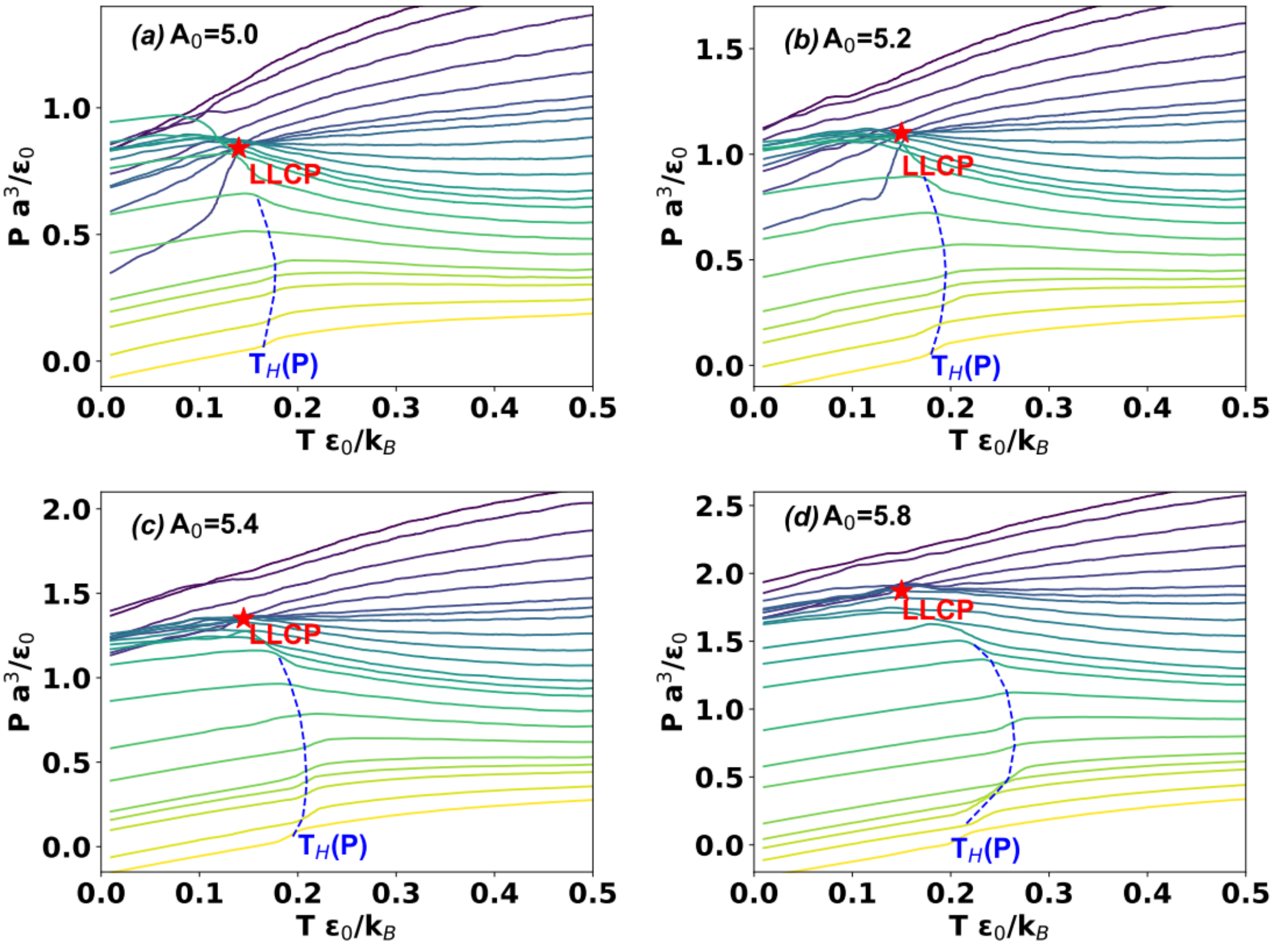

**Figure. S12. P-T phase diagrams of the FJ models with (a) $A_0$ = 5.0, (b) 5.2, (c) 5.4 and (d) 5.8.** Colored lines represent isochores obtained during the cooling at a rate $q_T = 10^{-4}\ \epsilon_0^{\frac{3}{2}}/m^{\frac{1}{2}}k_B a$. For the softer interactions ($A_0$ = 5.0 − 5.4), a liquid-liquid critical point (LLCP) is observed. For the hard interactions ($A_0$ = 5.8), LLCP can not be observed in equilibrium states. At low pressure, spontaneous crystallization occurs (blue dashed line). The LLCP location obtained here is used only as an initial estimate.

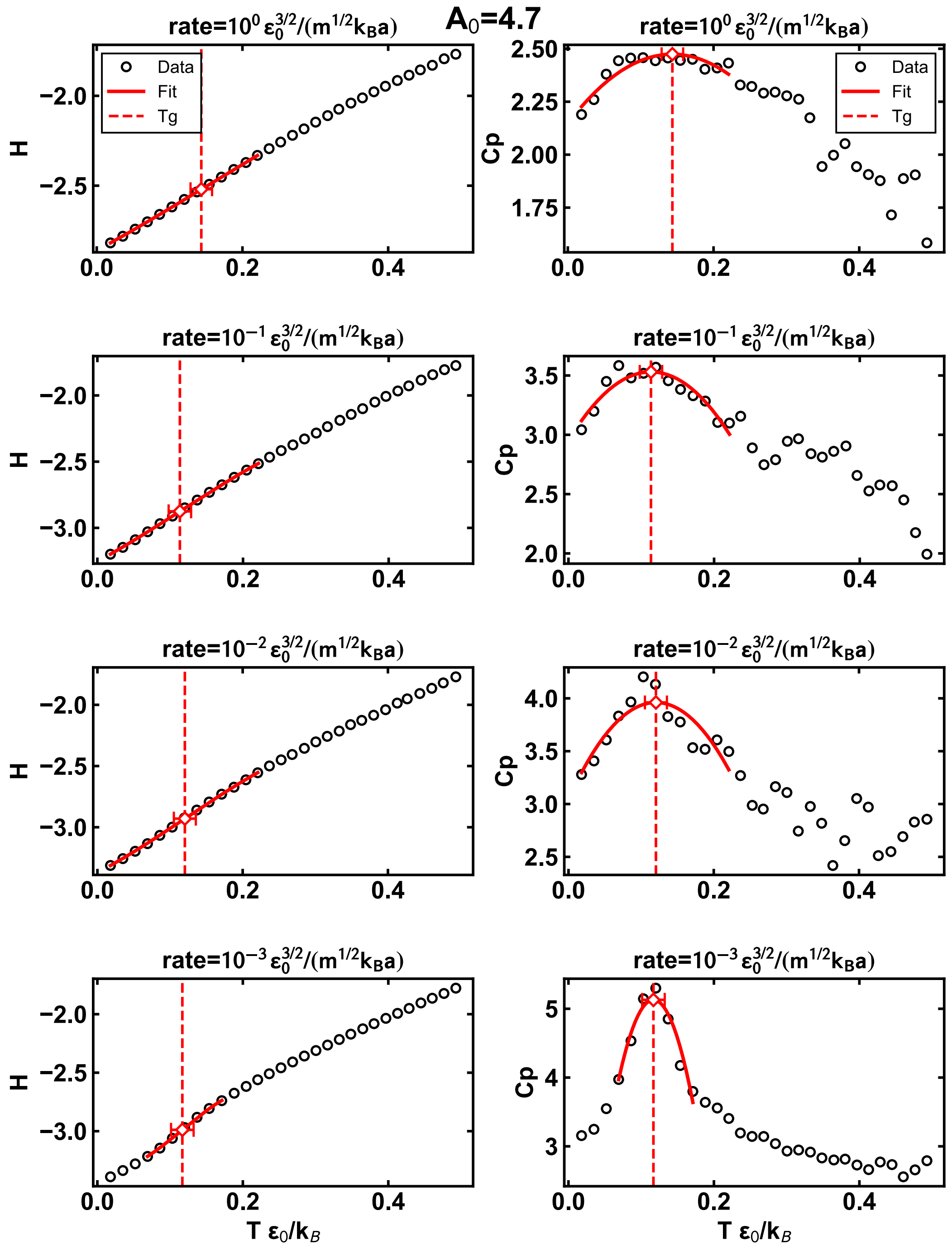

**Figure. S13. Determination of the glass-transition temperature ($T_g$) at critical pressure ($P_c$) for the $A_0 = 4.7$ model at different cooling rates.** The enthalpy ($H(T)$) was obtained from isobaric cooling simulations at cooling rates of ($q_T = 10^0 - 10^{-3}\epsilon_0^{3/2}/(m^{1/2}k_B a)$) (from top to bottom). The heat capacity was calculated from the temperature derivative of the enthalpy, ($C_P = \left(\frac{dH}{dT}\right)_P$). The glass-transition temperature

$(T_g)$ was determined from the maximum of $(C_P(T))$. Black circles represent the simulation data, red solid lines show the fits, and red dashed lines indicate the determined $(T_g)$.

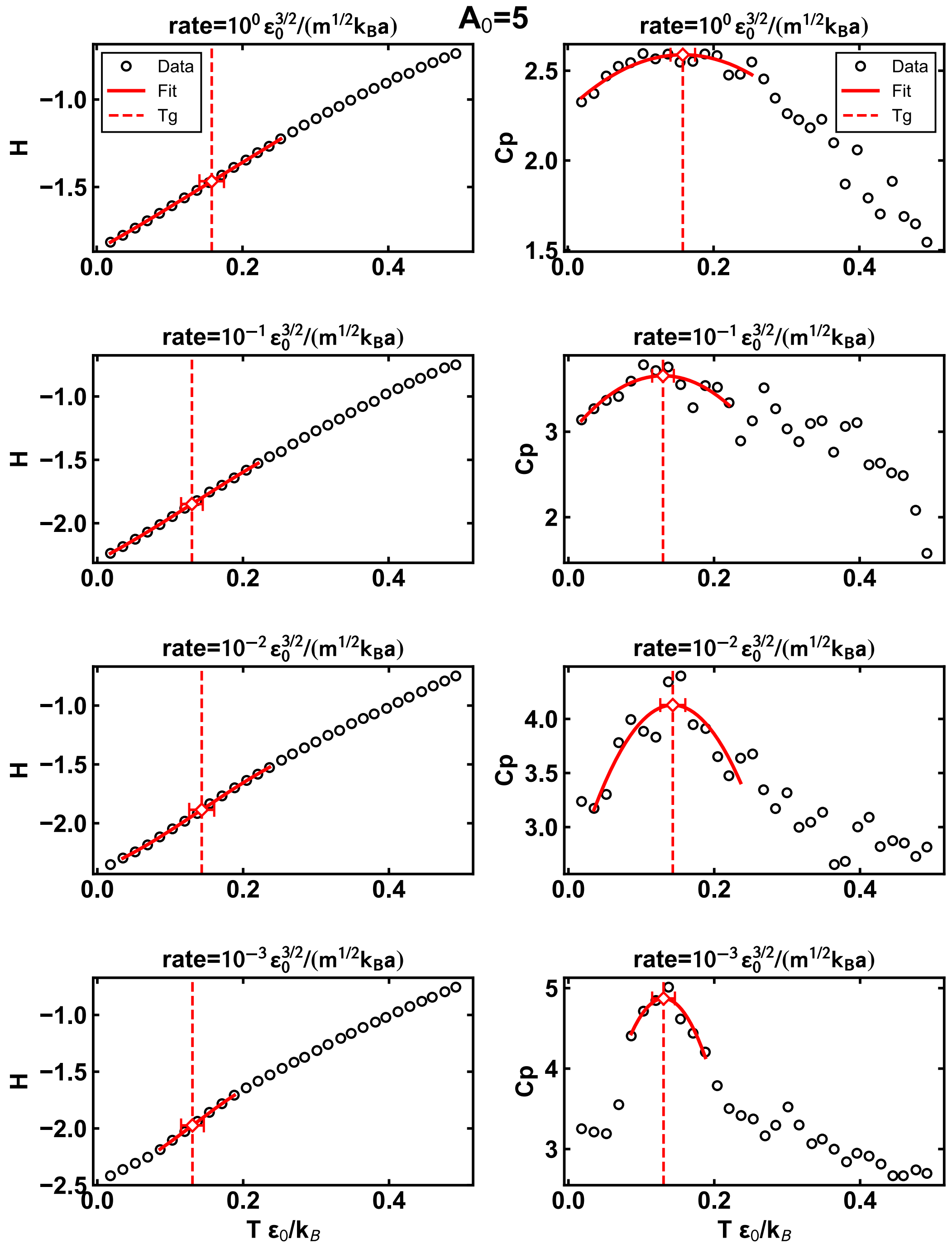

**Figure. S14. Determination of the glass-transition temperature ($\boldsymbol{T_g}$) at critical pressure ($\boldsymbol{P_c}$) for the $\boldsymbol{A_0 = 5.0}$ model at different cooling rates.** The enthalpy ($H(T)$) was obtained from isobaric cooling simulations at cooling rates of ($q_T = 10^0 - 10^{-3}\epsilon_0^{3/2}/(m^{1/2}k_B a)$) (from top to bottom). The heat capacity was calculated from the temperature derivative of the enthalpy, ($C_P = \left(\frac{dH}{dT}\right)_P$). The glass-transition temperature ($T_g$) was determined from the maximum of ($C_P(T)$). Black circles represent the simulation data, red solid lines show the fits, and red dashed lines indicate the determined ($T_g$).

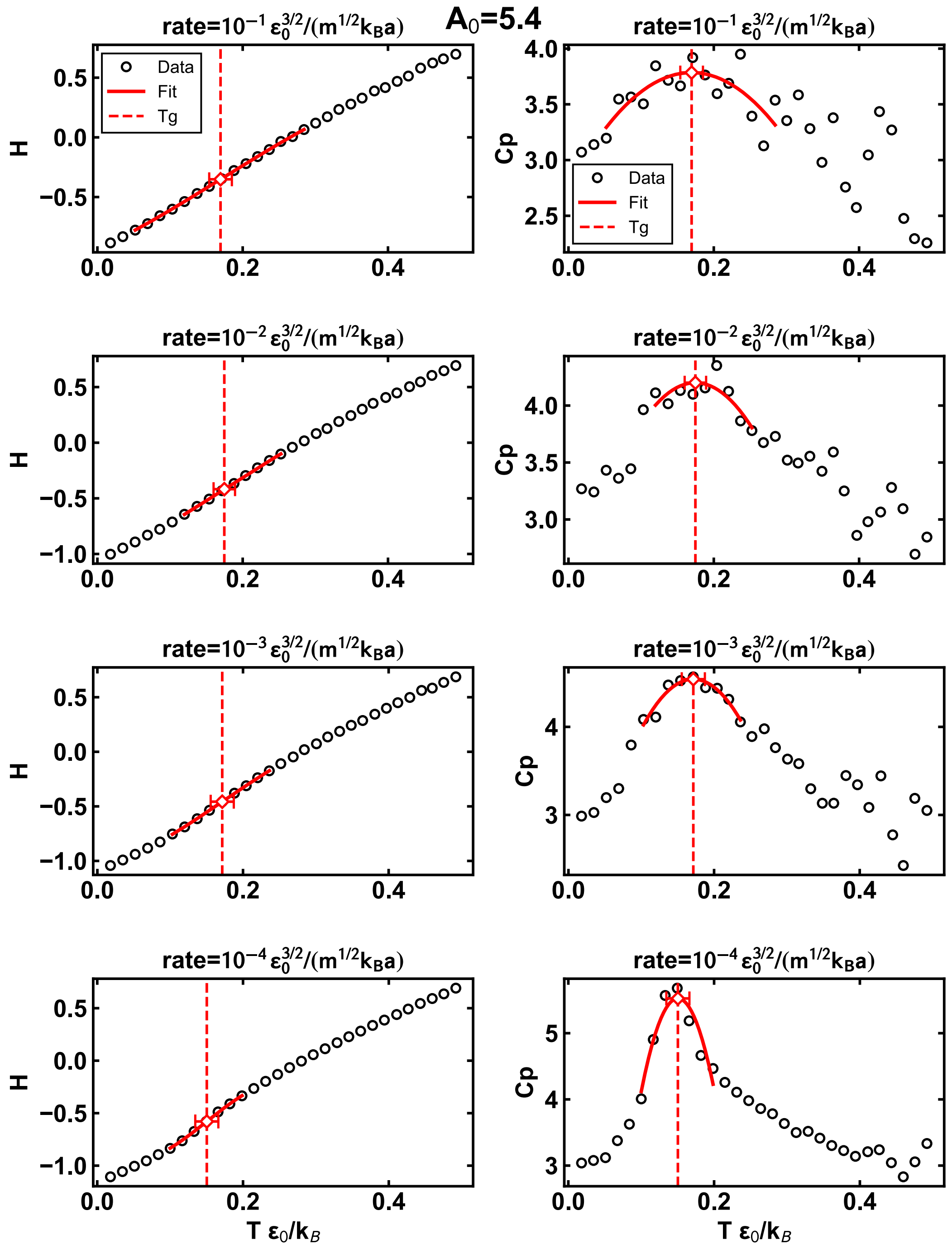

**Figure. S15. Determination of the glass-transition temperature ($T_g$) at critical pressure ($P_c$) for the $A_0 = 5.4$ model at different cooling rates.** The enthalpy ($H(T)$) was obtained from isobaric cooling simulations at cooling rates of ($q_T = 10^{-1} - 10^{-4}\epsilon_0^{3/2}/(m^{1/2}k_B a)$) (from top to bottom). The heat capacity was calculated from the temperature derivative of the enthalpy, ($C_P = \left(\frac{dH}{dT}\right)_P$). The glass-transition

temperature ($T_g$) was determined from the maximum of ($C_P(T)$). Black circles represent the simulation data, red solid lines show the fits, and red dashed lines indicate the determined ($T_g$).

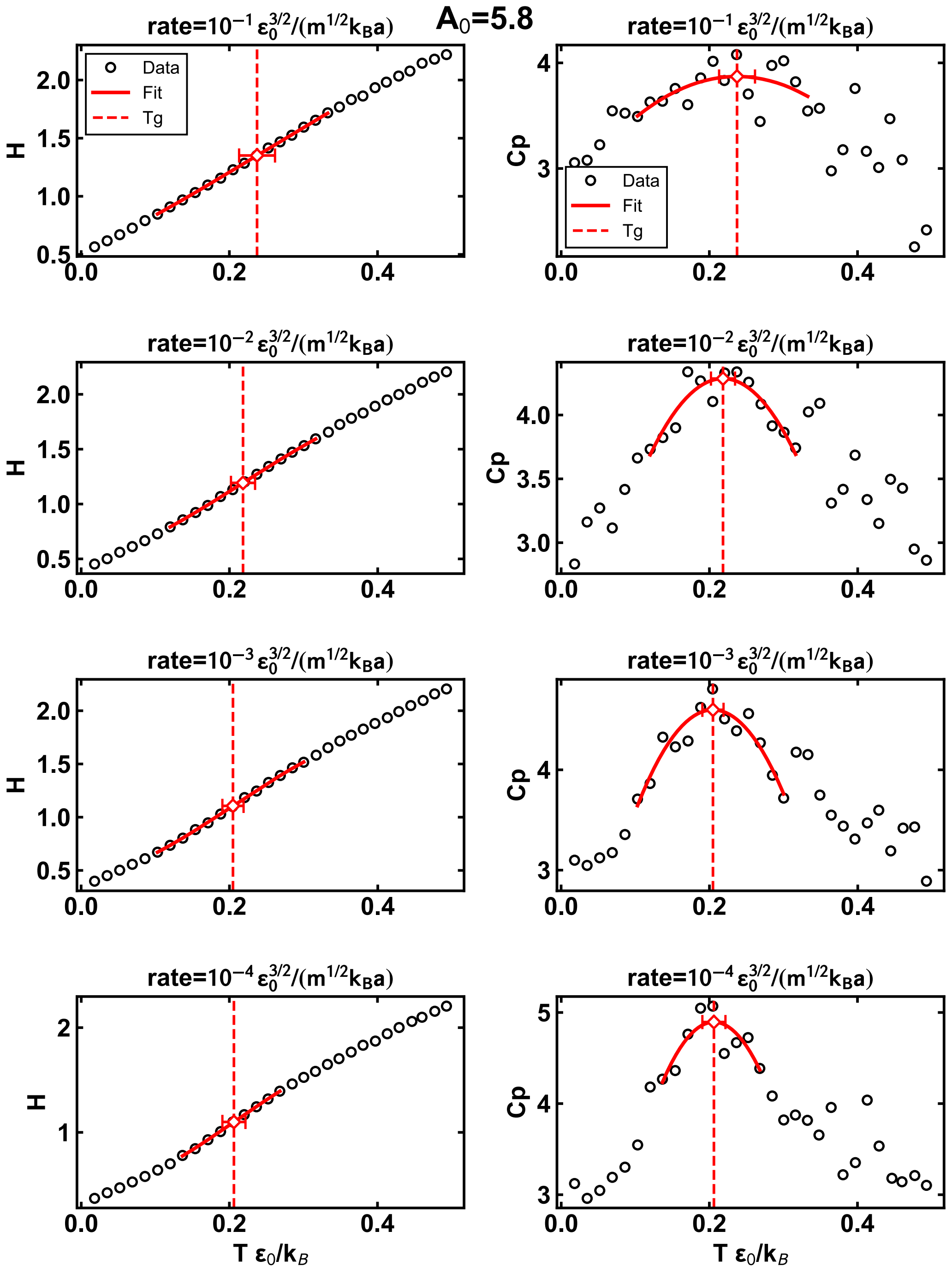

**Figure. S16. Determination of the glass-transition temperature ($\boldsymbol{T_g}$) at critical pressure ($\boldsymbol{P_c}$) for the $\boldsymbol{A_0 = 5.8}$ model at different cooling rates.** The enthalpy ($H(T)$) was obtained from isobaric cooling simulations at cooling rates of ($q_T = 10^{-1} - 10^{-4}\epsilon_0^{3/2}/(m^{1/2}k_B a)$) (from top to bottom). The heat capacity was calculated from the temperature derivative of the enthalpy, ($C_P = \left(\frac{dH}{dT}\right)_P$). The glass-transition temperature ($T_g$) was determined from the maximum of ($C_P(T)$). Black circles represent the simulation data, red solid lines show the fits, and red dashed lines indicate the determined ($T_g$).

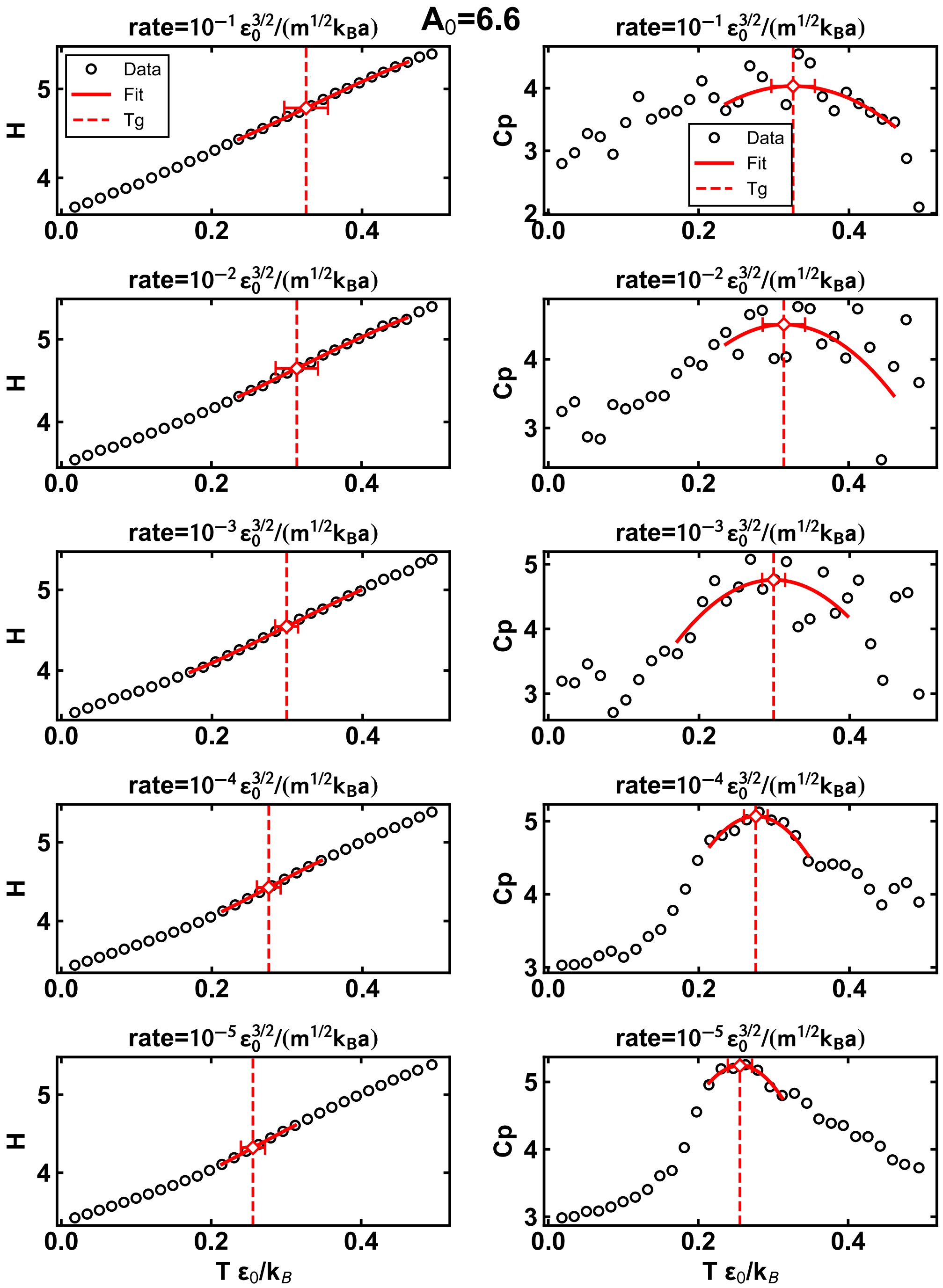

**Figure. S17. Determination of the glass-transition temperature ($T_g$) at critical pressure ($P_c$) for the $A_0 = 6.6$ model at different cooling rates.** The enthalpy ($H(T)$) was obtained from isobaric cooling simulations at cooling rates of ($q_T = 10^{-1} - 10^{-5}\epsilon_0^{3/2}/(m^{1/2}k_B a)$) (from top to bottom). The heat capacity was calculated from the temperature derivative of the enthalpy, ($C_P = \left(\frac{dH}{dT}\right)_P$). The glass-transition

temperature ($T_g$) was determined from the maximum of ($C_P(T)$). Black circles represent the simulation data, red solid lines show the fits, and red dashed lines indicate the determined ($T_g$).

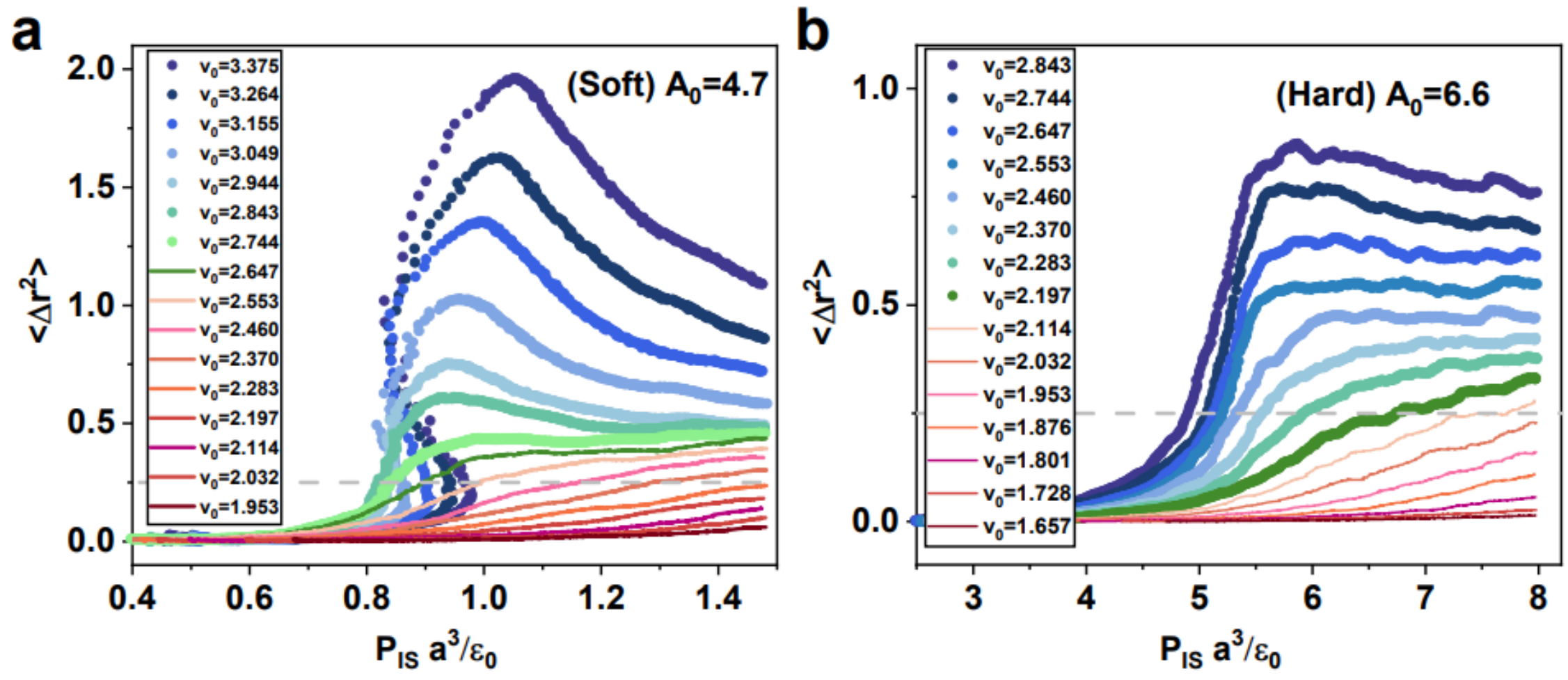

**Figure. S18.** The mean square displacement (MSD) of FJ liquid-derived ISs with (a) $A_0$ = 4.7 and (b) $A_0$ = 6.6 during compression, starting from inherent structures with different volumes. Circles mark LDL-derived ISs, and solid curves denote HDL-derived ISs.

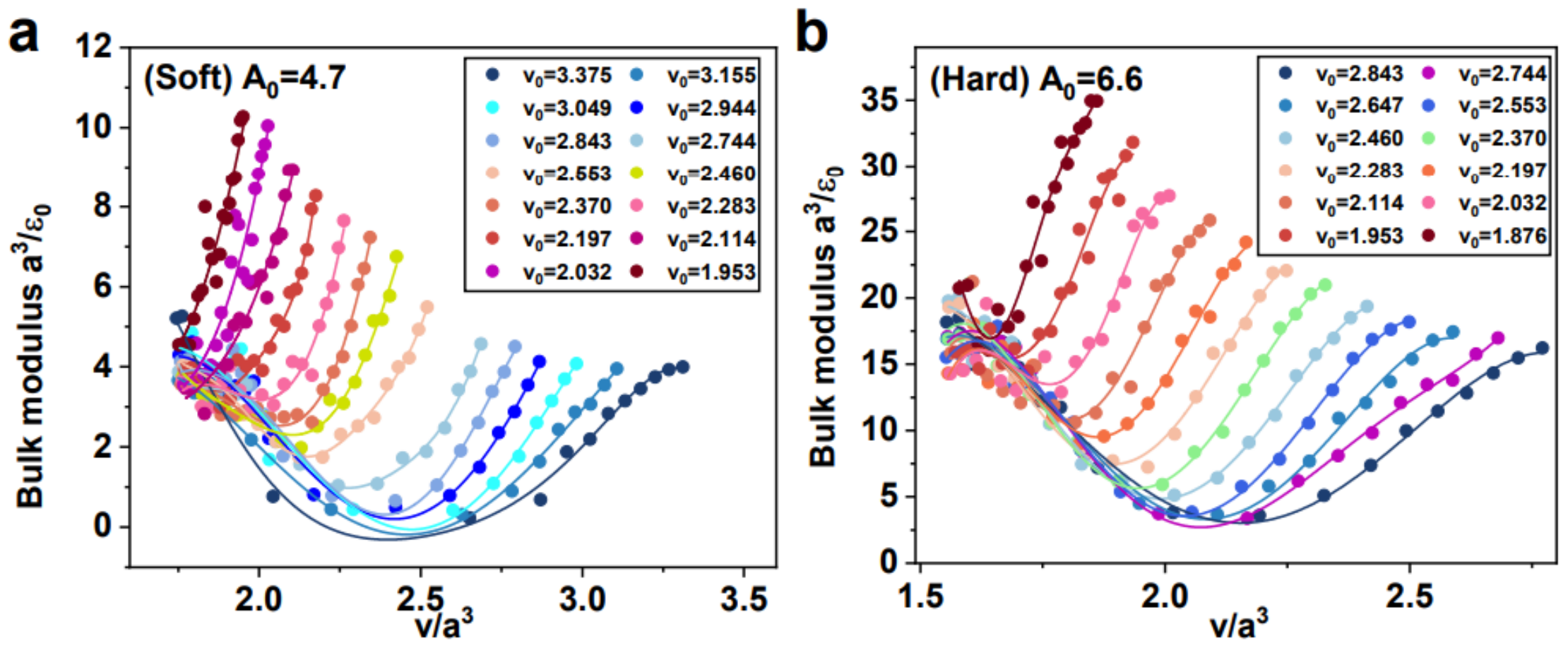

**Figure. S19.** The bulk modulus (B) of FJ liquid-derived ISs with (a) $A_0$ = 4.7 and (b) $A_0$ = 6.6 during pressure-induced transformation starting from inherent structures at different volumes. During sharp plastic deformation, $B$ reaches a minimum.

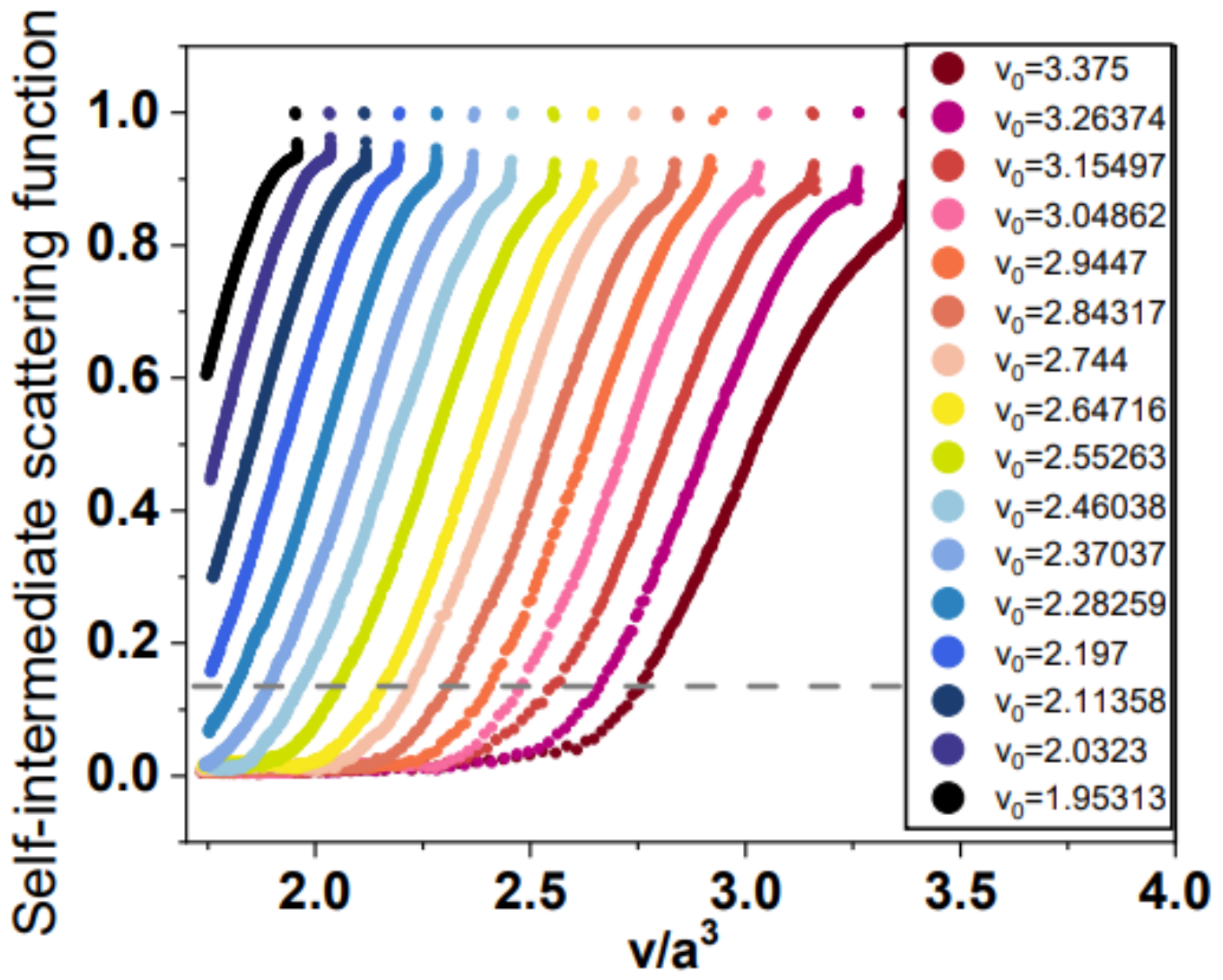

**Figure. S20.** Self-intermediate scattering function during compression of FJ liquid-derived ISs with $A_0$ = 4.7 for different volumes $v_0$ ranging from $1.95a^3$ to $3.38\ a^3$. The wave vector $k_0$ corresponds to the first peak of the structure factor S(k). At the crossover point, $F \approx e^{-2}$ .

# REFERENCES

1. N. Giovambattista, H. E. Stanley, F. Sciortino, Potential-energy landscape study of the amorphous-amorphous transformation in H2O. *Physical Review Letters* **91** (2003).
2. N. Giovambattista, F. Sciortino, F. W. Starr, P. H. Poole, Potential energy landscape of the apparent first-order phase transition between low-density and high-density amorphous ice. *Journal of Chemical Physics* **145** (2016).
3. P. H. Handle, F. Sciortino, N. Giovambattista, Glass polymorphism in TIP4P/2005 water: A description based on the potential energy landscape formalism. *Journal of Chemical Physics* **150** (2019).
4. Y. Z. Liu, G. Sun, L. M. Xu, Glass polyamorphism in gallium: Two amorphous solid states and their transformation on the potential energy landscape. *Journal of Chemical Physics* **154** (2021).